\documentclass[english,aps,prx,superscriptaddress,twocolumn]{revtex4}
\usepackage[T1]{fontenc}
\usepackage{color}
\usepackage{babel}
\usepackage{array}
\usepackage{verbatim}
\usepackage{refstyle}
\usepackage{amsmath}
\usepackage{amssymb}
\usepackage{graphicx}
\usepackage{esint}
\usepackage[unicode=true,pdfusetitle,
 bookmarks=true,bookmarksnumbered=false,bookmarksopen=false,
 breaklinks=false,pdfborder={0 0 1},backref=false,colorlinks=true]
 {hyperref}
\hypersetup{
 citecolor=blue}

\makeatletter

\providecommand{\tabularnewline}{\\}
\RS@ifundefined{subsecref}
  {\newref{subsec}{name = \RSsectxt}}
  {}
\RS@ifundefined{thmref}
  {\def\RSthmtxt{theorem~}\newref{thm}{name = \RSthmtxt}}
  {}
\RS@ifundefined{lemref}
  {\def\RSlemtxt{lemma~}\newref{lem}{name = \RSlemtxt}}
  {}

\@ifundefined{textcolor}{}
{%
 \definecolor{BLACK}{gray}{0}
 \definecolor{WHITE}{gray}{1}
 \definecolor{RED}{rgb}{1,0,0}
 \definecolor{GREEN}{rgb}{0,1,0}
 \definecolor{BLUE}{rgb}{0,0,1}
 \definecolor{CYAN}{cmyk}{1,0,0,0}
 \definecolor{MAGENTA}{cmyk}{0,1,0,0}
 \definecolor{YELLOW}{cmyk}{0,0,1,0}
}

\usepackage{prettyref}
\newrefformat{tab}{Table\,\ref{#1}}
\newrefformat{fig}{Figure\,\ref{#1}}
\newrefformat{eq}{Eq.\,\textup{(\ref{#1})}}

\makeatother

\begin{document}

\title{Geometrodynamics of electrons in a crystal under position and time
dependent deformation}

\author{Liang Dong}

\affiliation{International Center for Quantum Materials, Peking University, Beijing
100871, China}

\affiliation{Department of Physics, The University of Texas at Austin, Austin,
Texas 78712, USA}

\author{Qian Niu}
\email{niuphysics@utexas.edu}

\selectlanguage{english}%

\affiliation{Department of Physics, The University of Texas at Austin, Austin,
Texas 78712, USA}

\affiliation{International Center for Quantum Materials, Peking University, Beijing
100871, China}
\begin{abstract}
Semiclassical dynamics of Bloch electrons in a crystal under slowly
varying deformation is developed in the geometric language of a lattice
bundle. Berry curvatures and gradients of energy are introduced in
terms of lattice covariant derivatives, with the corresponding connections
given by the gradient and rate of strain. A number of physical effects
are discussed: an effective post-Newtonian gravity at band bottom,
polarization induced by spatial gradient of strain, orbital magnetization
induced by strain rate, and electron energy stress tensor.
\end{abstract}
\maketitle

\section{INTRODUCTION}

Semiclassical dynamics of Bloch electrons was developed in the early
days of solid state physics to give an intuitive picture of electron
motion in ionic background in a crystal. Combined with the Boltzmann
equation, many equilibrium and transport phenomena are well described
\cite{equationsofmotion}. The theory can also be quantized \cite{Peierlssubstitution}
to describe quantum energy levels, such as Wannier-Stark ladder in
a constant electric field and Landau levels in a constant magnetic
field \cite{equationsofmotion}. Much later, Berry phases \cite{Berrysphase}
were found to play an important role in the semiclassical equations
of motion in the form of Berry curvatures \cite{xiao2010berry}. It
accounts for quite a few phenomena such as the quantum Hall effect
\cite{TKNN}, the intrinsic anomalous Hall effect \cite{macdonalanomaloushall}
and charge pumping \cite{chargepumping}. The description of electric
polarization \cite{polarization} and orbital magnetization \cite{DOS}
in crystals are also closely related to the notion of Berry phase
.

The wave-packet method developed by Sundaram and Niu provides a systematic
way to derive the semiclassical dynamics in perturbed crystals \cite{sundaram}
with various types of Berry curvatures appearing in the equations
of motion. In their work, besides the effect of electromagnetic field,
they studied the crystal deformation. Some interesting results are
found such as dragging effect due to lattice motion and real space
Berry phase associated with a dislocation. Their theory is based on
the displacement field of ions, and the wave-packet method can easily
account for the effects to first order in spatial and time derivatives
of the displacement field, i.e the strain and lattice velocity. However,
it is very difficult to extend to the next order of derivatives, such
as strain gradient, lattice acceleration and strain rate in this formalism,
which are related to phenomena such as flexoelectricity \cite{Tagantsev1985,flexoelectrictensor}
and viscosity \cite{viscositylandau}. 

In this paper, we advocate another picture by viewing the whole crystal
under deformation as a lattice bundle. For a given time and about
each spatial point, a locally periodic lattice can be identified,
which well describes the distribution of ions in a region smaller
than the length scale of the variation of periodicity. The local lattice
also moves rigidly at the average velocity of ions around the position
point. By identifying the local lattice at each space and time point,
we have a bundle of locally periodic lattices on spacetime manifold.
Each local lattice contains information to the arbitrary order of
strain (defined relative to some reference lattice). Strain gradient,
strain rate and lattice acceleration is manifested in the difference
between local lattices. In this sense, periodic crystal can be viewed
as a special case where all local lattices are identical.

Based on lattice bundle picture, we have two geometrical structures,
which are extended phase space and Hilbert bundle \cite{hilbertbundle}.
Noticing each local lattice gives rise to a Brillouin zone, all the
Brillouin zones together with the spacetime make up the extended phase
space $(\boldsymbol{k};\boldsymbol{x},t)$, where crystal momentum
$\boldsymbol{\boldsymbol{k}}$ denote points in the Brillouin zone
of the local lattice at spacetime point $(\boldsymbol{x},t)$. However,
in a crystal under deformation, the sizes of local Brillouin zones
are different in general. This is in contrast to an ideal crystal
where all local Brillouin zones are the same and can be viewed as
a single one. This special geometry brings up the question about how
to express the electron equations of motion in phase space. Particularly,
we are faced with the difficulty of defining the change rate of crystal
momentum $\boldsymbol{k}$ since at different time slices crystal
momentums belong to different Brillouin zones. 

Moreover, the definition of Berry connections in terms of spacetime
parameter is also problematic. The naive idea from previous experience
is to define, for example, $A_{t}=i\langle u|\partial_{t}u\rangle$
using the eigenstates of the $\boldsymbol{k}$ dependent Hamiltonian
$\hat{H}(\boldsymbol{k};\boldsymbol{x},t)$ given by the local lattice.
However, $|\partial_{t}u\rangle$ involves the difference between
two Bloch functions of different periodicities which is not a periodic
function in general and gives rise to ill defined results. In fact,
this problem involves the second geometrical structure we mentioned
earlier called lattice Hilbert bundle. Noticing all the eigenstates
of Hamiltonian $\hat{H}(\boldsymbol{k};\boldsymbol{x},t)$ form a
complete basis for periodic functions thus give rise to a local Hilbert
space defined at a particular extended phase space point $(\boldsymbol{k};\boldsymbol{x},t)$,
the lattice Hilbert bundle appears as the collection of all local
Hilbert spaces together with the extended phase space. Similarly,
local Hilbert spaces at different spacetime points are distinct in
periodicities inherited from the corresponding local lattices. So
the definition of Berry connection actually involves comparison between
states in different Hilbert spaces. 

To resolve the aforementioned problem, with the same spirit in differential
geometry \cite{fecko2006differential}, we introduce the concept of
lattice covariant derivative to take the place of partial derivatives
in comparing local quantities such as crystal momentum, Bloch functions
and band energy. It gives mathematically and physically reasonable
results as shown later in this paper. With lattice covariant derivative,
we achieve our main result in this paper: the equations of motion
of electrons accurate to the first order of strain gradient, strain
rate and lattice acceleration. The results are expressed in terms
of the Bloch functions given by the local periodic lattice, which
are solvable numerically. A few interesting effects are discussed
based our lattice covariant formalism. First neglecting the Berry
phase effects, we show the similarity between the electron dynamics
at band bottom to that of a test particle in post-Newtonian gravity.
A equivalent metric tensor is identified in terms of the band structure
and lattice deformation. Then we focus on the Berry phase related
effects. For a band insulator at zero temperature, we calculate the
current induced by deformation. Particularly, we identify the polarization
contribution and give a explicit expression for the proper piezoelectric
constant \cite{Vanderbiltproperpolarization} in terms of Berry curvatures
expressed with lattice covariant derivative. Then we discuss the Chern-Simons
part of the strain gradient induced polarization and strain rate induced
orbital magnetization. Finally, for spatial homogeneous case we discuss
the electron stress energy tensor and its responses to ionic velocity
gradient and acceleration. 

This paper is organized as follows. Section \ref{sec:LATTICE-BUNDLE-PICTURE}
introduces the basic idea of lattice bundle picture and clears the
mathematical notion. In Section \ref{sec:LATTICE-COVARIANT-PHASE},
we discuss the special geometry of phase space. The equations of motion
without Berry phase effect are discussed in comparison to the gravitational
effect. In Section \ref{sec:LATTICE-COVARIANT-FORMULATION}, we discuss
the lattice covariant Berry curvatures and their related effects.
Subsection \ref{subsec:Applications} concludes the paper with aforementioned
applications. 

\section{LATTICE BUNDLE PICTURE\label{sec:LATTICE-BUNDLE-PICTURE}}

\subsection{Local lattice}

In this paper, our main results are expressed in lab frame $(\boldsymbol{x},t)$,
which is a Cartesian coordinate representing Minkovski spacetime.
$\boldsymbol{x}$ is the position coordinate which has three components
denoted by $\{i,j,k\}$ running at $1,2,3$. $t$ is the time coordinate.
Compound notion like $x^{u,\nu,\xi}$ is also used in this paper where
$u,\nu,\xi=0,1,2,3$ to include the time component represented by
$0$. 

The description of an ideal crystal in lab frame is given by its Bravais
lattice, which is a set of lattice points periodically aligned in
space. Each lattice point may represent several ions and has a integer
label $\{\boldsymbol{l}\}$. Its displacement vector from the lab
frame origin is denoted as $\{\boldsymbol{R}_{\boldsymbol{l}}=l^{\alpha}\boldsymbol{a}_{\alpha}+\boldsymbol{u}\}$.
Here we use $\{\alpha,\beta,\gamma\}$ running at $1,2,3$ to denote
crystalline directions. $\boldsymbol{l}$ is short for $\{l^{\alpha}\}$
which has three components. $\{\boldsymbol{a}_{\alpha}\}$ are the
three primitive lattice vectors and $\boldsymbol{u}$ is the position
of the zeroth lattice point. Einstein summation rule is applied through
out this paper. 

In the lattice bundle picture, local lattices are expressed in the
same way as ideal lattices only with $\{\boldsymbol{a}_{\alpha}(\boldsymbol{x},t),\boldsymbol{u}(\boldsymbol{x},t)\}$
becoming position and time dependent vector fields denoting the property
at a particular spacetime point $(\boldsymbol{x},t)$. Reciprocal
lattice vector fields $\{\boldsymbol{b}^{\alpha}(\boldsymbol{x},t)\}$
are defined through the relation $\boldsymbol{b}^{\alpha}\cdot\boldsymbol{a}_{\beta}=\delta_{\beta}^{\alpha}$,
where the ``crystallographer's definition'' is used and $\delta_{\beta}^{\alpha}$
is the Kronecker delta function. We use bold symbol to denote tensors.
Normal symbol with the same letter denotes their components. For example,
lattice vectors are written as $\boldsymbol{a}_{\alpha}=(a_{\alpha}^{1},a_{\alpha}^{2},a_{\alpha}^{3})$
and reciprocal lattice vectors as $\boldsymbol{b}_{i}^{\alpha}=(b_{1}^{\alpha},b_{2}^{\alpha},b_{3}^{\alpha})$.
Vectors and covectors are indicated by their upper and lower indices
of their components respectively. Thus the orthogonal relation written
in components reads $b_{i}^{\alpha}a_{\beta}^{i}=\delta_{\beta}^{\alpha}$.
If we only consider the three-dimensional space, $\{\boldsymbol{a}_{\alpha}(\boldsymbol{x},t)\}$
provides a basis for real space vector fields e.g. electron velocity
and $\{\boldsymbol{b}^{\alpha}(\boldsymbol{x},t)\}$ provides a basis
for covector fields e.g electron momentum. The completeness relation
for this basis reads $b_{j}^{\alpha}a_{\alpha}^{i}=\delta_{j}^{i}$.

\begin{figure}[!t]
\includegraphics[scale=0.25]{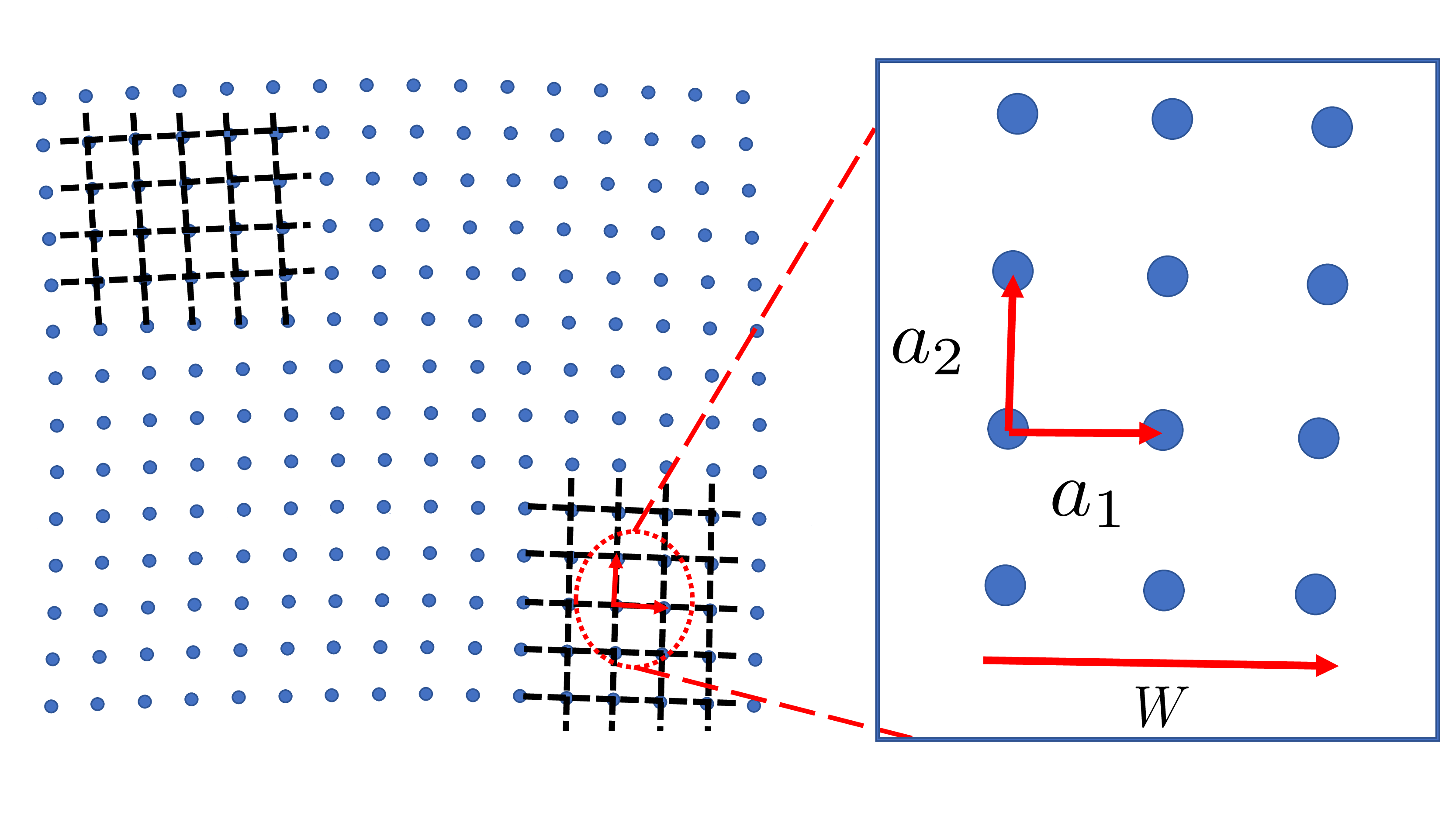}

\caption{A schematic picture of how to identify the local lattices in a deforming
crystal. Blue circle represents the lattice points of a deforming
crystal. Dashed lines represents the crystalline lines of the fictitious
periodic local lattice. As can be seen, the local distribution of
lattice points are well described by the local lattice only with deviation
far away from local point. The right panel is the zooming in picture.
The lattice vector of a local lattice is given by the relative displacement
between lattice points and located in the middle denoted by the black
dots on the arrows. And a local lattice moves rigidly at velocity
$\boldsymbol{W}$. }
\end{figure}

On the other hand, we know a crystal under deformation is described
by the position of its lattice points $\{\boldsymbol{R}_{\boldsymbol{l}}(t)\}$.
Thus we need to establish the relation between $\{\boldsymbol{a}_{\alpha}(\boldsymbol{x},t),\boldsymbol{u}(\boldsymbol{x},t)\}$
and $\{\boldsymbol{R}_{\boldsymbol{l}}(t)\}$. This relation is schematically
shown in figure (1) and is given by the following formula
\begin{align}
\boldsymbol{a}_{\alpha}(\frac{\boldsymbol{R}_{\boldsymbol{l}}+\boldsymbol{R}_{\boldsymbol{l}+1_{\alpha}}}{2},t)= & \boldsymbol{R}_{\boldsymbol{l}+1_{\alpha}}-\boldsymbol{R}_{\boldsymbol{l}},\label{eq:1-1}
\end{align}
where $1_{\alpha}$ means one increment in the $\alpha th$ crystalline
direction. To achieve the continuous lattice vector fields $\{\boldsymbol{a}_{\alpha}(\boldsymbol{x},t)\}$,
some interpolation procedure should be applied. A detailed discussion
is given in Appendix \ref{sec:Lattice-frame}. We should notice that
once the lattice vector fields are given, they can determine the total
crystal up to a rigid body displacement since given the position of
a particular lattice point, Eq. (\ref{eq:1-1}) can be applied repeatedly
to recover all the lattice points. The rigid body center position
is described by the field $\boldsymbol{u}(\boldsymbol{x},t)$ however
is insignificant due to the translational invariance of the Minkovski
spacetime. Thus $\boldsymbol{u}(\boldsymbol{x},t)$ will not be discussed. 

However, due to the time-dependence of the problem, it is convenient
to introduce a velocity field $\boldsymbol{W}(\boldsymbol{x},t)$
to describe the motion of local lattices. It is determined by the
property that
\begin{align}
\boldsymbol{W}(\boldsymbol{R}_{l},t)=\dot{\boldsymbol{R}}_{l}(t),\label{eq:4}
\end{align}
where again interpolation procedure should applied to achieve a continuous
field. Then the four fields $\{\boldsymbol{a}_{\alpha}(\boldsymbol{x},t),\boldsymbol{W}(\boldsymbol{x},t)\}$
with $\alpha=1,2,3$ define at each spacetime point $(\boldsymbol{x},t)$
a periodic lattice of periodicity $\{\boldsymbol{a}_{\alpha}(\boldsymbol{x},t)\}$
moving rigidly at velocity $\boldsymbol{W}(\boldsymbol{x},t)$ as
observed in lab frame. Thus we have a lattice bundle over the spacetime. 

It is worth to point out that the above argument is always applicable
to cases where the primitive unit cell only contains one ion. For
multi-ion cases, we need to check whether there exists a one-to-one
correspondence between the ionic positions in a unit cell and the
instantaneous positions of Bravais lattice points i.e all degrees
of freedom is contained in lattice points. Examples where such correspondence
exists are deformation caused by a particular acoustic phonon branch,
adiabatic case where ions are always relaxed to their minimal energy
position at each time slice or crystals satisfying clamped-ion approximation
where at each time the distribution of ions within a unit-cell follows
the instantaneous strain of the unit-cell (although there can be internal
strain contribution \cite{Tagantsevflexo}). Exceptions are cases
where deformation is caused by optical phonon or there is internal
strain. 

Each local lattice automatically gives rise to a local ``static''
Hamiltonian as

\begin{align}
\hat{H}(\boldsymbol{k};\boldsymbol{x},t)= & \frac{1}{2m_{e}}(\frac{1}{i}\frac{\partial}{\partial\boldsymbol{r}}+\boldsymbol{k})^{2}+V(\{l^{\alpha}a_{\alpha}(\boldsymbol{x},t)-\boldsymbol{r}\}).\label{7}
\end{align}
$\boldsymbol{r}$ is the real space representation. $\boldsymbol{k}$
is the crystal momentum. Both $\boldsymbol{r}$ and $\boldsymbol{k}$
should be viewed as local quantity as well. The domain of $\boldsymbol{r}$
is chosen to be the first Wigner\textendash Seitz cell of the local
lattice. And the domain of $\boldsymbol{k}$ is chosen to be the first
Brillouin zone. The detailed derivation to achieve this Hamiltonian
is given in Appendix \ref{sec:Lattice-frame} to \ref{sec:Lagrangian-of-wave-packet}
and the relation between $\boldsymbol{r}$ and $(\boldsymbol{x},t)$
is given by Eq. (\ref{eq:96}). 

The above Hamiltonian has translational symmetry and is solvable in
principle. Its eigenstates are the periodic part of the Bloch functions
denoted by $u_{n}(\boldsymbol{r},\boldsymbol{k};\boldsymbol{x},t)$,
where $n$ is the band index. We will call $u_{n}(\boldsymbol{r},\boldsymbol{k};\boldsymbol{x},t)$
Bloch functions in this paper just for simplicity. The corresponding
eigenvalue are band energy denoted by $\varepsilon_{n}(\boldsymbol{r},\boldsymbol{k};\boldsymbol{x},t)$.
In the following discussion, we assume those eigensolutions are given
from first principle calculations and all our results will be given
based on them. 

Here are a few comments about this local Hamiltonian. In the crystal
potential term, information about the center position of local lattice
is discarded as the zeroth ion is always located at $\boldsymbol{r}=0$.
However, as pointed out earlier, this won't cause trouble since the
lattice vector fields contain the structure information up to a rigid
body displacement. Also the local lattice velocity $\boldsymbol{W}(\boldsymbol{x},t)$
is absent. This is because the rigid body motion of local lattice
merely change the physical meaning of crystal momentum and band energy
to those relative to ions. Rather than being discussed in the Hamiltonian
level, we will introduce the effect of velocity directly in the semiclassical
equations of motion in \ref{subsec:Electron-equations-of}. Again,
this procedure can be rigorously proved in Appendix \ref{sec:Lagrangian-of-wave-packet}. 

Also to make sure the above local Hamiltonian gives a good approximation
to the real lattice Hamiltonian, we must assume the crystal potential
at a given position is mainly determined by ions within some spatial
length scale and the change of lattice vectors is negligible within
this spatial scale. Also, because we have applied adiabatic approximation,
where the instantaneous Bloch function and band energy are used, the
variation time scale should also be small compared to the local energy
band gap. However, for polar materials, there does exist a non-local
contribution to the crystal potential due to polarization. Although
this part of contribution is not the focus of this paper, we expect
the polarization contribution to crystal potential can be accounted
for by combining the static Poisson equation self-consistently with
our formula of polarization given in Subsec. \ref{subsec:Applications}.
Then our theory takes care of the local potential part in this complete
approach. 

\subsection{Lattice connection as strain rate and gradient}

The major motivation of this work is to study the effect of inhomogeneity
i.e lattice acceleration, strain gradient and strain rate on electron
semiclassical dynamics, which are described by the spacetime derivative
of the fields $\{\boldsymbol{a}_{\alpha}(\boldsymbol{x},t),\boldsymbol{W}(\boldsymbol{x},t)\}$.
In formulating the theory, we find it is more convenient to define
a quantity that is directly related to inhomogeneity called lattice
connection. 

Consider the lattice vectors change $\delta\boldsymbol{a}_{\alpha}=\partial_{\mu}\boldsymbol{a}_{\alpha}(\boldsymbol{x},t)dx^{\mu}$
given by a small increment in position and time $dx^{\mu}$, we can
define the lattice connection to encode this variation as

\begin{align}
\delta\boldsymbol{a}_{\alpha}(\boldsymbol{x},t)= & \boldsymbol{\Gamma}_{\mu}(\boldsymbol{x},t)\cdot\boldsymbol{a}_{\alpha}(\boldsymbol{x},t)dx^{\mu},\label{eq:11}
\end{align}
where $\boldsymbol{\Gamma}_{\mu}(\boldsymbol{x},t)$ is the lattice
connection and can be viewed as second rank tensor with its element
denoted by $\{\Gamma_{j\mu}^{i}\}$. In components, the above equation
reads $\delta a_{\alpha}^{i}=\Gamma_{j\mu}^{i}a_{\alpha}^{j}dx^{\mu}$.
If we define the infinitesimal strain as $\delta\boldsymbol{s}=dx^{\mu}\boldsymbol{\Gamma}_{\mu}$,
it simply reads $\delta\boldsymbol{a}_{\alpha}=\delta\boldsymbol{s}\cdot\boldsymbol{a}_{\alpha}$,
which is the strain induced lattice vector change. Thus the physical
meaning of lattice connection $\boldsymbol{\Gamma}_{\mu}$ is just
the gradient $\mu=1,2,3$ and rate $\mu=0$ of the unsymmetrized strain
tensor. The antisymmetric part between the upper and first lower index
is the relative rotation between local lattices and the symmetric
part is the relative symmetrized strain. Particularly, the unit-cell
volume change is described by $\Gamma_{i\mu}^{i}$. Since we choose
local lattice as reference, strain is no longer present in our theory
instead strain gradient denoted by lattice connection gives the leading
order correction. Multiplying $\boldsymbol{b}^{\alpha}$ on both sides
of Eq. (\ref{eq:11}) and summing over $\alpha$, we can have a explicit
expression of lattice connection as

\begin{align}
\Gamma_{j\mu}^{i}= & b_{j}^{\alpha}\partial_{\mu}a_{\alpha}^{i}.\label{eq:5}
\end{align}

Lattice connection represents how local lattices are connected together
to form the total lattice structure. If in a local region the deformation
is elastic i.e can be continuously deformed to an periodic crystal,
the gradient of the fields $\{\boldsymbol{a}_{\alpha}(\boldsymbol{x},t),\boldsymbol{W}(\boldsymbol{x},t)\}$
are not independent, which is directly manifested on the property
of lattice connection. As seen from Figure \ref{fig:2.-Elastic-condition},
the four adjacent local lattice vectors forming a closed quadrilateral
leads to the conclusion that
\begin{align}
(\boldsymbol{a}_{\alpha}\cdot\boldsymbol{\partial})\boldsymbol{a}_{\beta}-(\boldsymbol{a}_{\beta}\cdot\boldsymbol{\partial})\boldsymbol{a}_{\alpha}= & 0,\label{5}
\end{align}
where the second order derivatives of $\{\boldsymbol{a}_{\alpha}(\boldsymbol{x},t)\}$
are ignored. On the other hand, if we consider the relative velocity
between two adjacent lattice points, it can be expressed either as
the total time derivative of the lattice vector fields from Eq. (\ref{eq:1-1})
or as the gradient of velocity field from Eq. (\ref{eq:4}). Equating
both expressions leads to the relation that
\begin{align}
\partial_{t}|_{x}\boldsymbol{a}_{\alpha}+(\boldsymbol{W}\cdot\boldsymbol{\partial})\boldsymbol{a}_{\alpha}=(\boldsymbol{a}_{\alpha}\cdot\boldsymbol{\partial})\boldsymbol{W} & ,\label{6}
\end{align}
where again second order derivatives of $\boldsymbol{W}(\boldsymbol{x},t)$
are neglected. Both Eq. (\ref{5}) and (\ref{6}) can be reformulated
in terms of lattice connection as
\begin{align}
\Gamma_{ij}^{k}-\Gamma_{ji}^{k}= & 0,\label{eq:12-1}\\
\Gamma_{i0}^{k}+\Gamma_{ji}^{k}W^{j}= & \partial_{i}W^{k}.\label{eq:13-1}
\end{align}
The first relation is the torsion free condition for a connection
form in a coordinate basis, which tells that the two lower indexes
of lattice connection are symmetric. Eq. (\ref{eq:13-1}) shows that
the strain rate experienced by ions the left hand side equals the
gradient of velocity field the right hand side, %
{} where the first term on the left hand side is the strain rate observed
at a fixed position. From the above expression, we see that the angular
velocity field given by $\boldsymbol{\omega}=\boldsymbol{\partial}\times\boldsymbol{W}$
is also related to lattice connection as $\omega^{k}=\varepsilon^{ijk}(\Gamma_{i0}^{j}+\Gamma_{li}^{j}W^{l})$
with $\varepsilon^{ijk}$ the Levi-Civita symbol. Thus angular velocity
is just the antisymmetric part of the strain rate experienced by ions.
However, we should notice that the lattice acceleration field $\boldsymbol{a}$
is not directly related to lattice connection, which is given by $\boldsymbol{a}=(\boldsymbol{W}\cdot\boldsymbol{\partial})\boldsymbol{W}+\partial_{t}\boldsymbol{W}$. 

Besides its geometrical meaning, lattice connection also gives rise
to a important gradient correction to local Hamiltonian (\ref{7}).
This gradient correction term is for the electron wave-packet, which
is localized in real space with center position $\boldsymbol{x}$
and expressed as the superposition of the local Bloch states of local
Hamiltonian (\ref{7}) \cite{sundaram}. For such a wave-packet state,
the gradient correction reads: 

\begin{align}
\Delta H(\boldsymbol{r};\boldsymbol{x},t)= & \frac{1}{2}\Gamma_{ni}^{m}\hat{O}_{m}^{ni}+\frac{1}{2}\Gamma_{ni}^{m}[(r-r_{c})^{i}\hat{D}_{m}^{n}+C.C].\label{eq:19-1}
\end{align}
where $\boldsymbol{r}_{c}$ is the expectation value of operator $\boldsymbol{r}$
under wave-packet state. The derivation is given in Appendix \ref{sec:Gradient-expansion-to}
and Appendix \ref{sec:Schrodinger-equation-in}. The first term comes
from the difference between local lattice and real lattice away from
the particular point where local lattice resides. The second term
comes from the difference between local lattices. Because wave-packet
states have finite sizes usually as large as several unit-cells, electrons
can feel the influence from adjacent local lattices as well. Both
terms are proportional to the spatial part of lattice connection.

$\hat{O}_{m}^{ni}=\sum_{l}\frac{\partial V(\{\tilde{\boldsymbol{R}}_{l}-\boldsymbol{r}\})}{\partial\tilde{R}_{l}^{m}}(\tilde{R}_{l}-r)^{n}(\tilde{R}_{l}-r)^{i}$
is a periodic operator respect to $\boldsymbol{r}$, where $\{\tilde{R}_{l}^{i}(\boldsymbol{x},t)=l^{\alpha}a_{\alpha}^{i}(\boldsymbol{x},t)\}$
are the local lattice points. The second term breaks the translational
symmetry due to the factor $\{\boldsymbol{r}-\boldsymbol{r}_{c}\}$.
However, it still has a well-defined expectation value under wave-packet
state due to localization in real space. $\hat{D}_{m}^{n}=\mathcal{V}_{m}^{n}-\frac{1}{m_{e}}(\frac{1}{i}\frac{\partial}{\partial r^{m}}+k_{m})(\frac{1}{i}\frac{\partial}{\partial r^{n}}+k_{n})$
is the deformation potential operator \cite{PhysRev.121.720} with
$\mathcal{V}_{m}^{n}=\sum_{l}\frac{\partial V(\{r-\tilde{R}_{l}(x,t)\})}{\partial\tilde{R}_{l}^{m}}(r-\tilde{R}_{l})^{n}$.
$\mathcal{V}_{m}^{n}$ has the same form as assumed in the rigid ion
model \cite{rigidionmodel} and automatically vanishes in the deformable
ion model. Actually, follow the argument in this paper \ref{subsec:Energy-correction-and},
we believe this form of $\mathcal{V}_{n}^{m}$ is rather general as
long as there is a one to one correspondence between all lattice points
and the crystal potential. Deformation potential operator is a second
rank symmetric tensor operator with respect to lattice symmetries.
Its expression in momentum space is discussed in Appendix G, which
might be more useful in first principle calculations. 

Local Hamiltonian (\ref{7}) together with gradient correction (\ref{eq:19-1})
give the total Hamiltonian of electron wave-packet in the first order
of strain gradient or lattice connection. Although in principle gradient
correction also modifies local Bloch states, the eigenstates of local
Hamiltonian (\ref{7}) are enough to achieve the equations of motion
up to first order. 

\subsection{Characterizing line defects in the lattice bundle picture\label{subsec:characterizing-line-defects}}

\begin{figure}[!t]
\includegraphics[scale=0.24]{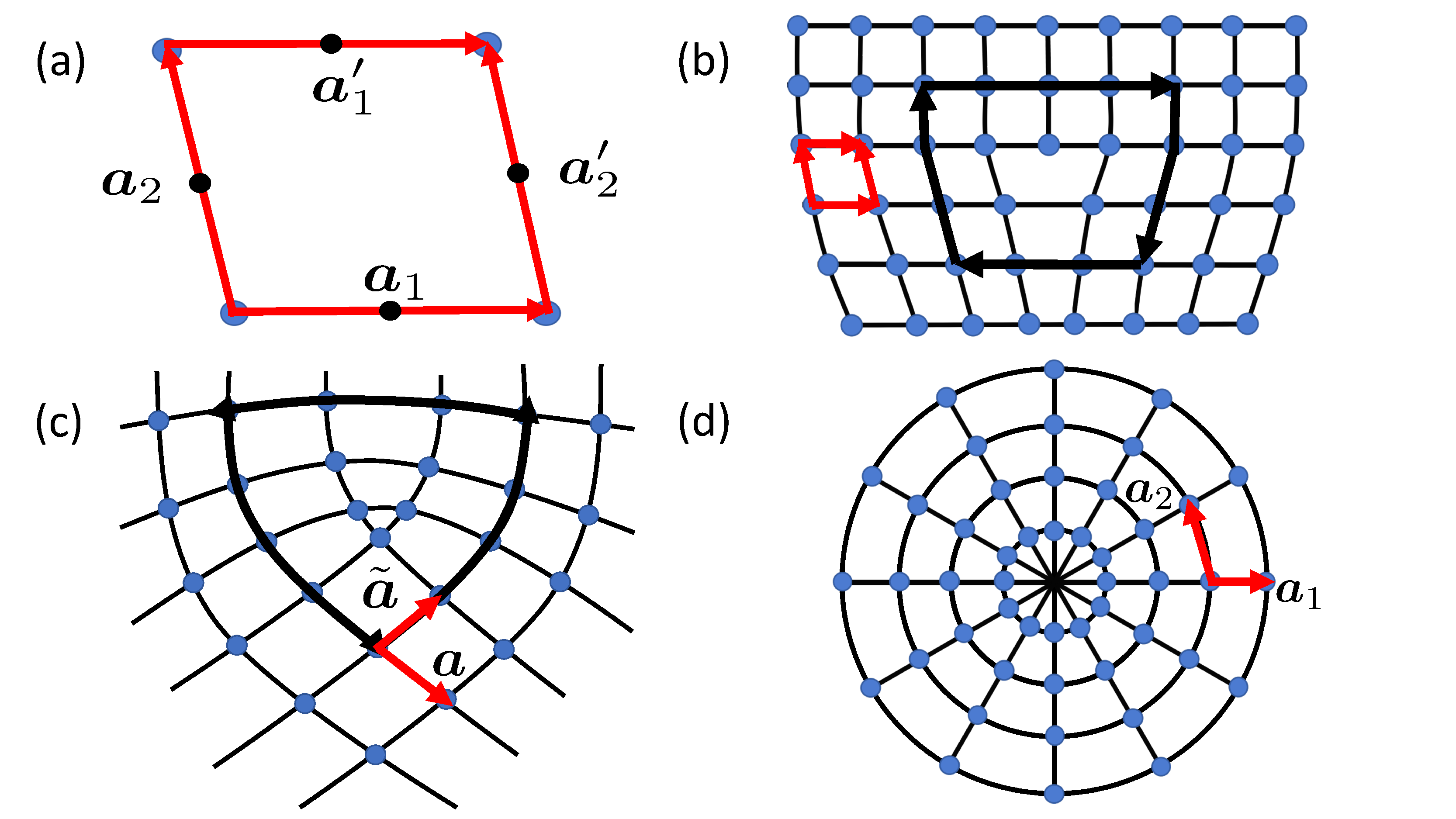}

\caption{(a) Elastic condition of lattice vectors. The dots on the four corners
represent lattice points, which determine the local lattice vectors
denoted by the four edges of the quadrilateral. The dot in the middle
of each edge denotes where these lattice vectors resides. According
to Eq. (\ref{eq:1-1}), it is straight forward to derive the condition
for the four lattice vectors to form a closed quadrilateral. (b) Schematics
of dislocation in a square lattice. Along the closed trajectory denoted
by bold line, if we count the change of lattice label, we always find
a unit mismatch in the crystalline direction. (c) Schematics of a
disclination in square lattice. Along the closed trajectory denoted
by bold line, we find the lattice vector continuously change from
$a$ to $\tilde{a}$ a shown in the picture (d) A special case of
disclination in a rectangular lattice, which is characterized by the
$2\pi$ rotation of lattice vector along the defect center. In the
region far enough from the center, locally we have a rectangular lattice,
whose two lattice vectors $\boldsymbol{a}_{1}=c_{1}\boldsymbol{\theta},\boldsymbol{a}_{2}=c_{2}\boldsymbol{r}$
are along the angle direction and radius direction respectively in
a polar coordinate. The disclination located at the origin is characterized
by the rotation of lattice vectors by $2\pi$ along the complete circle.
\label{fig:2.-Elastic-condition}}
\end{figure}

In this subsection, we discuss how to describe line defects within
lattice bundle picture. Although our theory is limited in regions
where locally deformation is slowly varying and elastic, the topology
of line defects can still be described by the loops enclosing the
defect line \cite{homotopytheory}. Here we consider the cases of
dislocation and disinclination. 

These two kinds of defect correspond to the redundant freedom to describe
a Bravais lattice. For an ideal crystal described by $\{\boldsymbol{a}_{\alpha},\boldsymbol{u}\}$,
we can change the lattice labels by some integer $\boldsymbol{Z}=\{Z^{\alpha}\}$
with $\boldsymbol{u}$ fixed, which gives the same lattice. This is
associated with dislocation. If we travel along the loop enclosing
dislocation line, after going back to the initial point, we find the
lattice labels are changed by some integer. Also, we may choose other
crystalline directions $\tilde{\boldsymbol{a}}_{\alpha}=U_{\alpha}^{\beta}\boldsymbol{a}_{\beta}$,
such that $\{\tilde{\boldsymbol{a}}_{\alpha},\boldsymbol{u}\}$ gives
the same lattice. This degree of freedom is related to disclination.
In the lattice bundle picture, if we keep track of the local lattice
vectors change along the loop enclosing disclination, after returning
to the starting point, we end up with another set of equivalent local
lattice vectors. Dislocation and disclination are topological in the
sense that $\boldsymbol{Z}$ is a integer vector and $\boldsymbol{U}$
is a integer matrix.

The above argument is schematically shown in figure \ref{fig:2.-Elastic-condition}(b)
and (c). Here, we show how they can be described mathematically. First,
for dislocation, Eq. (\ref{5}) tells us that that primitive lattice
vectors equals to the position change of lattice points per integer
label increment. This can be written in a discrete form as
\begin{align}
\delta x^{i}= & a_{\alpha}^{i}(\boldsymbol{x}+\frac{\delta\boldsymbol{x}}{2},t)\delta l^{\alpha}.\label{eq:13-2}
\end{align}
The above expression relates the change in $\boldsymbol{x}$ space
and $\boldsymbol{l}$ space. With it, we can calculate the change
of $\boldsymbol{x}$ or $\boldsymbol{l}$ along some trajectory. If
the trajectory is a loop in $\boldsymbol{l}$ space, the displacement
in $\boldsymbol{x}$ space gives the Burgers vector. On the other
hand, if the trajectory forms a loop in $\boldsymbol{x}$ space, the
total change of $\boldsymbol{l}$ gives the mismatch of lattice label
$\boldsymbol{Z}$ mentioned before. Here we adopt the latter perspective.
Considering a loop in $\boldsymbol{x}$ space far away from and enclosing
dislocation line, the change of lattice label can be written as a
integral as 
\begin{align}
\oint_{C}b_{i}^{\alpha}(\boldsymbol{x})dx^{i}= & Z^{\alpha},\label{eq:14}
\end{align}
where the integer on the right hand side describes the topological
``charge'' of the dislocation. According to whether the crystalline
direction $\boldsymbol{Z}$ is perpendicular or parallel to the plane
of the loop, we can characterize the dislocation either as screw dislocation,
edge dislocation or mixed type. For example, in figure \ref{fig:2.-Elastic-condition}b,
we have a edge dislocation in a square lattice with $\boldsymbol{Z}=(1,0,0)$
denoting one associated with dislocation line.

Next we discuss disclination. Eq. (\ref{eq:11}) gives a formal description
of how local lattice vectors change in position and time. Given lattice
connection and the initial value of lattice vectors $\{\boldsymbol{a}_{\alpha}\}$,
this equation determines the final value $\{\tilde{\boldsymbol{a}}_{\alpha}\}$
along some trajectory. For a loop enclosing disclination line, in
general the initial and final value of lattice vectors are different.
They are related to each other as $\tilde{\boldsymbol{a}}_{\alpha}=\boldsymbol{U}\cdot\boldsymbol{a}_{\alpha}$
and the matrix $\boldsymbol{U}$ can be formally expressed as
\begin{align}
\boldsymbol{U}= & \mathcal{T}exp(\oint dx^{i}\boldsymbol{\Gamma}_{i}),\label{eq:16}
\end{align}
where $\mathcal{T}$ is the path ordering operator which is necessary
when the matrixes $\boldsymbol{\Gamma}_{i}(x,t)$ at different points
along the loop do not commute. Given lattice connection expressed
in lab frame basis $\{\Gamma_{ni}^{m}\}$ , $\boldsymbol{U}$ will
have the form of $\{U_{n}^{m}\}$. However, to see its topological
property, we express it in lattice vector basis as $U_{\alpha}^{\beta}=a_{\alpha}^{n}U_{n}^{m}b_{m}^{\beta}$.
Then the final and initial value of lattice vectors are related as
$\tilde{\boldsymbol{a}}_{\alpha}=U_{\alpha}^{\beta}\boldsymbol{a}_{\beta}$.
Because $\{\tilde{\boldsymbol{a}}_{\alpha}\}$ and $\{\boldsymbol{a}_{\alpha}\}$
represent the same local lattice, the matrix $U$ and its inverse
are both integer matrix with determinant $\pm1$. 

It is important to realize that in the presence of disinclination,
lattice vector fields are not globally well-defined. This restricts
our previous discussion only to a local region. To study the global
effect of disinclination, at least two sets of lattice vector fields
are needed. However, lattice connection is still a good quantity globally.
As seen from its expression (\ref{eq:5}), the summation over all
crystalline directions makes lattice connection single-valued even
in the presence of disclination. It is also worth to point out that
because of elastic condition (\ref{eq:12-1},\ref{eq:13-1}), locally
lattice connection is trivial in the sense it can be made zero by
a particular coordinate transformation. However, in the presence of
topological defect, lattice connection is no longer trivial globally. 

Here, we give an demonstration calculation for disclination shown
by figure \ref{fig:2.-Elastic-condition}(d). In polar coordinates
$\boldsymbol{a}_{1}=c_{1}\boldsymbol{\theta},\boldsymbol{a}_{2}=c_{2}\boldsymbol{r}$,
where $c_{1},c_{2}$ are constant representing the magnitude of lattice
vector. Then according to Eq. (\ref{eq:11}), we have $\Gamma_{x\theta}^{x}=\Gamma_{y\theta}^{y}=0$
and $\Gamma_{x\theta}^{y}=-\Gamma_{y\theta}^{x}=\frac{1}{R}$, where
$R$ is the radius of the circle we are considering enclosing the
disclination. The lattice connection can be treated as a matrix

\begin{align}
\Gamma_{\theta}= & \left(\begin{array}{cc}
0 & -\frac{1}{R}\\
\frac{1}{R} & 0
\end{array}\right),
\end{align}
which is just the generator of the SO(2) group multiplied by $\frac{1}{R}$.
Because lattice connection is constant along the path, the ordering
operator can be omitted and we have

\begin{align}
U= & exp\left(\begin{array}{cc}
0 & -2\pi\\
2\pi & 0
\end{array}\right)=\left(\begin{array}{cc}
\cos(2\pi) & -\sin(2\pi)\\
\sin(2\pi) & \cos(2\pi)
\end{array}\right),\label{eq:16-1}
\end{align}
which is the expected $2\pi$ rotation of the lattice vectors. 

\section{LATTICE COVARIANT PHASE SPACE\label{sec:LATTICE-COVARIANT-PHASE}}

\subsection{Phase space geometry}

\begin{figure}[!t]
\includegraphics[scale=0.24]{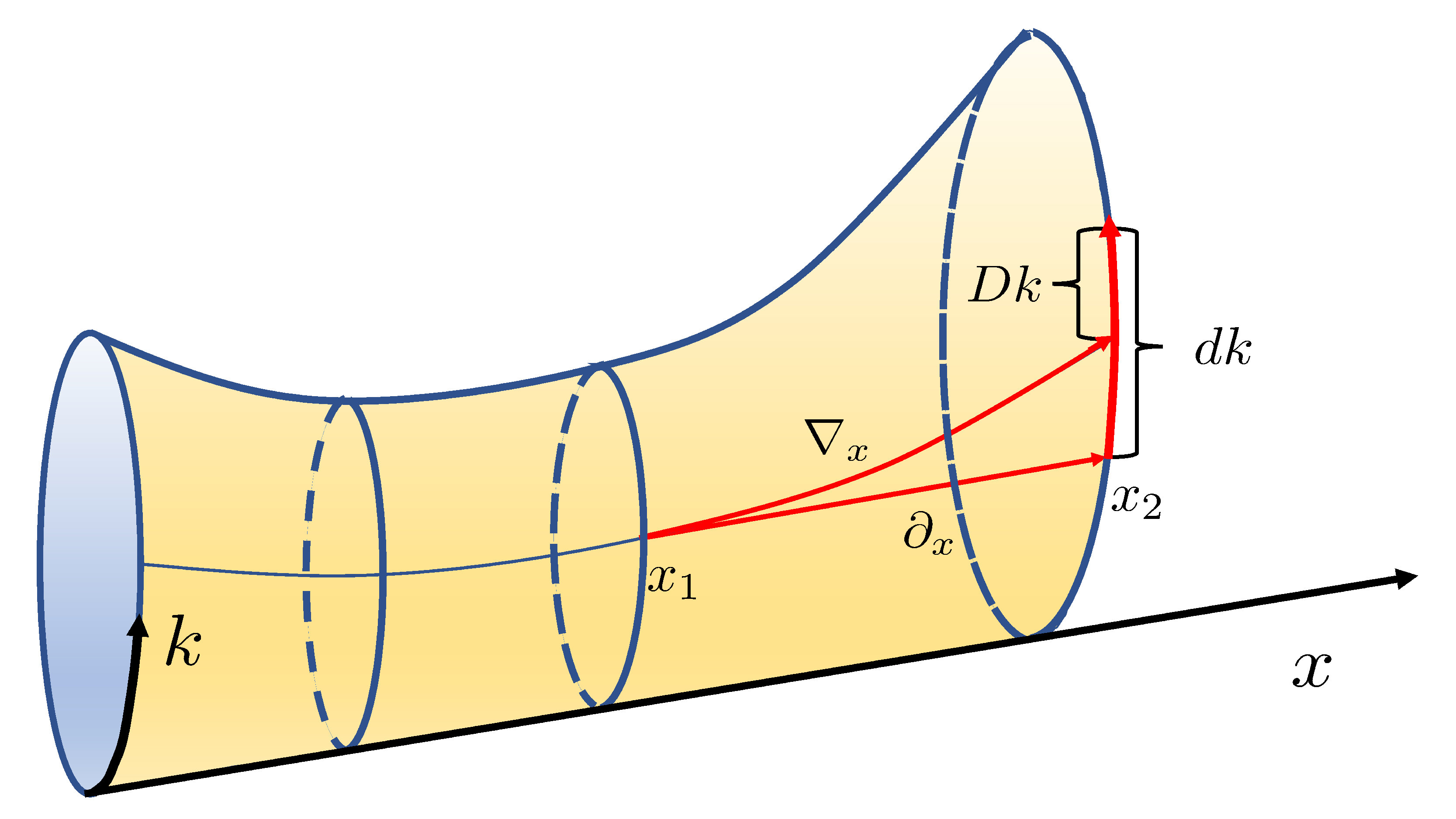}

\caption{\label{figure 3}The phase space of electrons in one-dimensional case.
The straight line at the bottom denotes the position space $\boldsymbol{x}$.
Local Brillouin zones are denoted by circles at each position. The
intersection between each circle and the straight line is the $\boldsymbol{k}=0$
point. The phase space is a ``tube'' with varying radius. To compare
two points in phase space, we show two alternatives. One is along
the path denoted by $\partial_{\boldsymbol{x}}$ where the value of
$\boldsymbol{k}$ is fixed then go along the circle by total change
of wave-vector $d\boldsymbol{k}$. Another path is go along the correspondence
line denoted by $\nabla_{\boldsymbol{x}}$ then go along the circle
by the mechanical change of wave-vector $D\boldsymbol{k}$.}
\end{figure}

At a given time $t$, the electron wave-packet state with center position
$\boldsymbol{x}$ and center wave-vector $\boldsymbol{k}$ locates
at the point $(\boldsymbol{k};\boldsymbol{x})$ in phase space. Phase
space is the base manifold for the semiclassical electron motion.
However, for the case of deforming crystals it takes a unusual geometry
comparing to periodic lattices as shown in figure \ref{figure 3},
which is the one-dimensional case. In the lattice bundle picture,
at given time $t$, at each position $\boldsymbol{x}$ the local lattice
gives rise to a local Brillouin zone according to its own periodicity.
All the Brillouin zones together with the position space $\boldsymbol{x}$
constitute the phase space. However, the shapes of local Brillouin
zones are different. Noticing the topology of Brillouin zone, we have
a bundle of smoothly varying toruses as the phase space. As shown
in the one-dimensional case, the phase space is a irregular tube. 

Mathematically, each point in a local Brillouin zone is labeled by
$\boldsymbol{\boldsymbol{k}}$ and we choose the domain of $\boldsymbol{k}$
as $[-\pi\boldsymbol{b}(\boldsymbol{x},t),\pi\boldsymbol{b}(\boldsymbol{x},t)]$,
with $-\pi\boldsymbol{b}$ and $\pi\boldsymbol{b}$ denoting the same
point. The position and time dependence of $\boldsymbol{b}(\boldsymbol{x},t)$
shows how the local Brillouin zones vary under deformation. Due to
this geometry, unlike periodic crystals, the meaning of wave-vector
$\boldsymbol{k}$ is incomplete without pointing out which local Brillouin
zone it belongs to. This property brings up the question about how
to compare wave-vectors in different local Brillouin zones. To answer
this question, a correspondence between local Brillouin zones is needed.
Thus we introduce the concept of correspondence curves in phase space.
Given a wave-vector $\boldsymbol{k}$ at some initial position point
$\boldsymbol{x}$, we move the wave-vector in real space while at
the same deform the wave-vector with local Brillouin zones. The infinitesimal
changes of wave-vector by a small shift in position is given by $\delta k_{m}=-\Gamma_{mi}^{n}k_{n}dx^{i}$. 

Wave-vectors on the same correspondence line are treated equivalently.
Thus it is useful to introduce a derivative operation $\nabla_{x^{\mu}}$
to encode this equivalence, which we call lattice covariant derivative.
Lattice covariant derivative is also crucial for the semiclassical
dynamics. For example, the band energy $\varepsilon(\boldsymbol{\boldsymbol{\boldsymbol{k}}},\boldsymbol{x},t)$
given by Hamiltonian (\ref{7}) is a time-dependent phase space function
and its derivative in position gives the ``force'' term responsible
for the acceleration of electrons. An important property of $\varepsilon$
is that it is periodic in $\boldsymbol{k}$ which makes it compatible
with the torus topology of Brillouin zone. Thus it is natural to require
that their lattice covariant derivatives are also periodic:

\begin{align}
\nabla_{x^{\mu}}f|_{(\boldsymbol{k}+2\pi\boldsymbol{b},\boldsymbol{x},t)}\equiv & \nabla_{x^{\mu}}f|_{(\boldsymbol{k};\boldsymbol{x},t)},
\end{align}
where $f$ is an arbitrary periodic function, $\nabla_{x^{\mu}}$
is the lattice covariant derivative and $\mu=0$ accounts for the
time derivative. It is easy to see that partial derivative with $\boldsymbol{k}$
fixed doesn't satisfy the above relation. To find the right derivative
operation, we consider the total differential change of an arbitrary
phase space $f(\boldsymbol{k},\boldsymbol{x},t)$: 
\begin{align}
df(\boldsymbol{k};\boldsymbol{x},t)= & [dx^{\mu}\nabla_{x^{\mu}}+Dk_{i}\partial_{k_{i}}]f,\label{eq:25}
\end{align}
with
\begin{align}
\nabla_{x^{\mu}}f= & (\partial_{x^{\mu}}-k_{l}\Gamma_{j\mu}^{l}\partial_{k_{j}})f,\label{eq:26}\\
Dk_{i}= & dk_{i}+k_{j}\Gamma_{i\mu}^{j}dx^{\mu},\label{eq:27}
\end{align}
where instead of using partial derivatives to express the total differential,
we rearrange the terms to write it in a lattice covariant form. $dx^{\mu}\nabla_{x^{\mu}}f$
is the differential change along the correspondence line and $Dk_{i}\partial_{k_{i}}$
is the change along $\boldsymbol{k}$, which is schematically shown
in figure \ref{figure 3}. 

Eq. (\ref{eq:26}) gives the desired lattice covariant derivative
operation. It is easy to check that the lattice covariant derivative
of $f$ is still periodic. As a bonus, we get another quantity $D\boldsymbol{k}$.
Noticing that $-\Gamma_{\mu}kdx^{\mu}$ is the geometrical change
of wave-vector due to deformation, $D\boldsymbol{k}$ is the total
change subtracting the geometrical change. So we call $D\boldsymbol{k}$
the mechanical change of wave-vectors. It is worth to point out that
the mechanical change of wave-vector defined this way is also periodic
in $\boldsymbol{k}$. 

When $f(\boldsymbol{k},\boldsymbol{x},t)=\varepsilon(\boldsymbol{k},\boldsymbol{x},t)$,
its lattice covariant derivative is related to the deformation potential
as
\begin{align}
\nabla_{x^{\mu}}\varepsilon(\boldsymbol{k};\boldsymbol{x},t)=D_{m}^{n}(\boldsymbol{k};\boldsymbol{x},t)\Gamma_{n\mu}^{m},\label{eq:28}
\end{align}
where $D_{m}^{n}\equiv(a_{\alpha}^{n}\partial_{a_{\alpha}^{m}}-k_{m}\partial_{k_{n}})\varepsilon$
is the deformation potential defined in the entire Brillouin zone\cite{deformationpotential}.
Here a trick has been used that when the position and time dependence
is through the lattice vectors, partial derivatives can be written
as $\partial_{x^{\mu}}=\Gamma_{n\mu}^{m}a_{\alpha}^{n}\partial_{a_{\alpha}^{m}}|_{k}$.
Usually the deformation potential is defined relative to a particular
reference crystal. Here we have a deformation potential tensor field
from all local lattices. 

\subsection{Electron equations of motion \label{subsec:Electron-equations-of}}

Next, we show how the equations of motion of electrons can be written
in a covariant form with the help of lattice covariant derivatives.
In this subsection, we neglect the Berry phase effect first. Without
Berry phase, the electron position $\boldsymbol{x}$ and wave-vector
$\boldsymbol{k}$ are a pair of canonical variables. Under single
band approximation, for a static deformed crystal, it is straight
forward to write down the equations of motion:
\begin{align}
\dot{\boldsymbol{x}}= & \partial_{\boldsymbol{k}}\varepsilon,\label{eq:25-1}\\
\dot{\boldsymbol{k}}= & -\partial_{\boldsymbol{x}}\varepsilon,\label{eq:26-1}
\end{align}
where $\varepsilon$ and $\boldsymbol{k}$ are the eigen energy and
eigenstate of the local Hamiltonian (\ref{7}). Although this form
is mathematically correct, the equations of motion are not compatible
with the phase space geometry mentioned before since $\partial_{\boldsymbol{x}}\varepsilon$
is not a periodic function in $\boldsymbol{k}$ and breaks the torus
topology of Brillouin zones. We can then rewrite the above equations
with lattice covariant derivatives as
\begin{align}
\dot{\boldsymbol{x}}= & \partial_{\boldsymbol{k}}\varepsilon,\label{eq:25-1-1}\\
D_{t}\boldsymbol{k}= & -\nabla_{\boldsymbol{x}}\varepsilon,\label{eq:26-1-1}
\end{align}
where not only every term in the above equations can be interpreted
as a quantity in phase space but also Eq. (\ref{eq:26-1-1}) acquires
new geometrical meaning that the mechanical change of wave-vector
is induced by the deformation potential force noticing Eq. (\ref{eq:28}). 

For the time-dependent case, the above equations of motion need to
be further modified. As pointed out earlier that the Hamiltonian is
written in the comoving frame of ions, thus $\varepsilon$ and $\boldsymbol{k}$
denotes the energy and crystal momentum relative to ions respectively.
Keeping this in mind, the equations of motion need to be revised in
two aspects. First, the energy dispersion represent the relative velocity
of electrons to ions thus the LHS of (\ref{eq:25-1-1}) should be
replaced by $\dot{\boldsymbol{x}}-\boldsymbol{W}$ instead. Second,
due to non-inertial motion of ions, inertial force should also contribute
to the change of the relative crystal momentum $\boldsymbol{k}$ in
Eq. (\ref{eq:26-1-1}). 

By adding those corrections, we achieve the equations of motion in
the most general case as
\begin{align}
D_{t}\boldsymbol{x}= & \partial_{\boldsymbol{k}}\varepsilon,\label{eq:30}\\
D_{t}\boldsymbol{k}= & -\nabla_{\boldsymbol{x}}\varepsilon+m_{e}D_{t}\boldsymbol{x}\times2\boldsymbol{\omega}-m_{e}\boldsymbol{a},\label{eq:31}
\end{align}
where $D_{t}\boldsymbol{x}\equiv\dot{\boldsymbol{x}}-\boldsymbol{W}$
is the relative velocity to ions. $m_{e}D_{t}\boldsymbol{x}\times2\boldsymbol{\omega}$
, $-m_{e}\boldsymbol{a}$ are the inertial forces due to lattice rotation
and acceleration respectively, where $\boldsymbol{\omega}=\frac{1}{2}\boldsymbol{\partial}\times\boldsymbol{W}$
is the angular velocity field and $\boldsymbol{a}=(\boldsymbol{W}\cdot\boldsymbol{\partial})\boldsymbol{W}+\partial_{t}\boldsymbol{W}$
is the acceleration field. The mechanical change rate of the wave-vector
is defined as $D_{t}\boldsymbol{k}\equiv\frac{D\boldsymbol{k}}{dt}$
and in components it reads:
\begin{align}
D_{t}k_{i}\equiv & (\dot{k}_{i}+\Gamma_{it}^{n}k_{n}+\dot{x}^{j}\Gamma_{ij}^{n}k_{n}).\label{eq:32}
\end{align}

The geometrical meaning of the above equations of motion becomes explicit
with the help of lattice covariant derivative. In fact if we compare
the equations of motion to those of a test particle moving in real
gravitational field \cite{michaelstone}, we can see they share a
lot of similarities. Eq. (\ref{eq:30}) is just the expression of
the covariant velocity if $\boldsymbol{W}$ is treated as $G^{oi}$
component of a metric with only the free particle energy dispersion
on the right hand side replaced by the band energy dispersion. $D_{t}\boldsymbol{k}$
has the form of the covariant derivative of the crystal momentum along
the electron trajectory in spacetime, which is the same as test particle
case. However, in real gravity the right hand side of Eq. (\ref{eq:31})
vanishes for spinless particles. One of the reasons for this distinction
is that our lattice connection is only for the spatial part of tangent
space of spacetime manifold while the Levi-Civita connection is for
the total tangent space. We expect a more complete analogy to gravity
can be made by considering the deformation of a Bloch-Floquet crystal\cite{gomez2013floquet}.
Another reason is that the deformation potential is not completely
geometrical in the sense different crystals have different form of
deformation potential. It is also worth to point out that the equations
of motion has the covariant property under Newtonian coordinate transformation.
Particularly, each term in Eq. (\ref{eq:30}) transforms like vector
and each term in Eq. (\ref{eq:31}) transforms as covector. This property
allows us to get the equations of motion in other coordinates related
by Newtonian coordinate transformation with time universal in all
coordinate choices. 

Chronologically, we derived the above equations of motion from the
following zeroth order Lagrangian:
\begin{align}
L_{e}^{0}= & -(\varepsilon(\boldsymbol{k};\boldsymbol{x},t)+\boldsymbol{W}\cdot\boldsymbol{k}+\frac{1}{2}m_{e}\boldsymbol{W}^{2})+(\boldsymbol{k}+m_{e}\boldsymbol{W})\cdot\dot{\boldsymbol{x}},\label{eq:22}
\end{align}
where the first order terms such as Berry connections and gradient
energy are discarded. This Lagrangian can be understood in terms of
its free electron limit, where $\varepsilon=\frac{k^{2}}{2m_{e}}$
and by defining the lab frame canonical momentum $\boldsymbol{p}\equiv\boldsymbol{k}+m_{e}\boldsymbol{W}$,
we have the expected free electron Lagrangian $L_{e}^{0}=p\dot{x}-\frac{p^{2}}{2m_{e}}$.
Direct variation of the Lagrangian (\ref{eq:22}) gives the following
equations of motion expressed in terms of partial derivatives
\begin{align}
\dot{\boldsymbol{x}}= & \partial_{\boldsymbol{k}}\varepsilon+\boldsymbol{W},\label{eq:23}\\
\dot{\boldsymbol{k}}= & -\partial_{\boldsymbol{x}}\varepsilon-\partial_{\boldsymbol{x}}\boldsymbol{W}\cdot\boldsymbol{k}+m_{e}(\dot{\boldsymbol{x}}-\boldsymbol{W})\times2\boldsymbol{\omega}-m_{e}\boldsymbol{a}.\label{eq:24}
\end{align}
However, as mentioned earlier the geometrical meaning of this form
is less obvious and cannot be interpreted as the equations of motion
in phase space. By using lattice covariant derivative and the local
elastic relation $\Gamma_{i0}^{k}+\Gamma_{ji}^{k}W^{j}=\partial_{i}W^{k}$
, they can be rewritten in the covariant form (\ref{eq:30}) and (\ref{eq:31}).

\subsection{Post-Newtonian gravity at band bottom}

Motivated by the similarity to gravitational effect, we study the
low energy dynamics around band extrema and find that the electron
dynamics is described by an effective post Newtonian gravity. Since
we only consider the deformation of crystals with spatial periodicity
and adopt the Newtonian point of view about time passing uniformly
regardless the deformation, it is unlikely to get a full analogy to
the four-dimensional gravity. However, it is reasonable to compare
to post-Newtonian gravity, which is the low energy and speed limit
of the complete gravitational theory. At band minimal, expanding local
energy to the second order of $\boldsymbol{k}$ and expressing the
electron wave-vector $\boldsymbol{k}$ in terms of $\dot{x}$, we
have the Lagrangian (\ref{eq:22}) as 
\begin{align}
\tilde{L}_{e}= & \frac{1}{2}m_{ij}^{*}(\boldsymbol{x},t)\dot{x}^{i}\dot{x}^{j}-(m_{ij}^{*}-m_{e}\delta_{ij})W^{j}(\boldsymbol{x},t)\dot{x}^{i}\nonumber \\
 & +\frac{1}{2}(m_{ij}^{*}-m_{e}\delta_{ij})W^{i}W^{j}-\Phi(\boldsymbol{x},t),\label{eq:33}
\end{align}
where $m_{ij}^{*}$ is the effective mass, $\Phi(\boldsymbol{x},t)$
is the energy at band extremals. It reduces to the Lagrangian of Newtonian
free particle when electron and lattice are decoupled. We compare
it with the Lagrangian of a electron in post Newtonian gravity. By
assuming both the velocities of the massive object generating gravity
and the test particle are small compared to the velocity of light
and keeping to the second order of velocity, the test particle's Lagrangian
reads
\begin{align}
\tilde{L}_{g}\approx & \frac{1}{2}m_{e}g_{ij}\dot{x}^{i}\dot{x}^{j}+m_{e}g_{0i}\dot{x}^{i}+m_{e}\phi,\label{eq:34}
\end{align}
where $g_{00}=-c^{2}+2\phi$ and $\phi$ is the Newton's gravitational
potential. The rest energy of electron is discarded. Direct comparison
of both Lagrangians leads to the equivalent metric tensor in lattice
as 
\begin{align}
\frac{m_{ij}^{*}}{m_{e}} & \sim g_{ij},\nonumber \\
-(\frac{m_{ij}^{*}}{m_{e}}-\delta_{ij})W^{j} & \sim g_{0i},\nonumber \\
\frac{1}{2}(\frac{m_{ij}^{*}}{m_{e}}-\delta_{ij})W^{i}W^{j}-\frac{\Phi}{m_{e}} & \sim\phi.\label{eq:35}
\end{align}
We emphasize that these effective metric is expressed in the lab frame.
The motion of electrons follows the geodesic equation given by the
above equivalent metric. The first relation between effective mass
tensor and the spatial part of the metric tensor is well known in
general solid state physics \cite{effctivemass}. The energy at band
bottom as a static potential serves as the Newton's gravitational
potential is also expectable. The new discovery here is the contribution
from ionic motion. Particularly the $g_{0i}$ component comes directly
from the ionic velocity. It acts as the vector potential of the gravitational
electro-magnetic field thus can be coupled to the energy magnetization
in the system. It also gives a dynamical contribution to the gravitational
potential. The factor $\frac{m_{ij}^{*}}{m_{e}}-\delta_{ij}$ which
is proportional to the difference between effective mass and bare
mass is a manifestation of the dragging effect. For free electrons
whose effective mass is just the bare mass this effect vanishes. For
effective mass larger and smaller than the bare mass, $g_{0i}$ just
have opposite effects. It is also interesting to study the inverse
effect of electrons on lattice dynamics in crystals which corresponds
to the effect of massive objects on the dynamics of gravitational
field.

\section{LATTICE COVARIANT FORMULATION OF BERRY PHASE EFFECTS\label{sec:LATTICE-COVARIANT-FORMULATION} }

In this section, we focus on the Berry phase effects. From previous
discussion, we know that evolution of a electron is described by its
trajectory in phase space with a special geometry. Under adiabatic
approximation, the electron Bloch functions will change adiabatically
along the trajectory. If the trajectory forms a loop, the initial
and final state of Bloch functions only differ by a phase term called
Berry phase. The corresponding Berry connection and Berry curvature
will modify the previous equations of motion (\ref{eq:30}, \ref{eq:31}). 

\subsection{Lattice covariant Berry connections and Berry curvatures}

\begin{table*}[!t]
\begin{tabular}{|c|>{\centering}p{6cm}|>{\centering}p{7cm}|}
\hline 
phase space quantities & examples & lattice covariant derivatives\tabularnewline
\hline 
scalar functions & $\varepsilon(\boldsymbol{k};\boldsymbol{x},t)$ & $\nabla_{x^{\mu}}\varepsilon\equiv\partial_{x^{\mu}}\varepsilon-\Gamma_{m\mu}^{n}k_{n}\frac{\partial\varepsilon}{\partial k_{m}}$\tabularnewline
\hline 
vectors & $A_{k}(\boldsymbol{k};\boldsymbol{x},t)$ & $\nabla_{x^{\mu}}A_{k_{i}}\equiv\partial_{x^{\mu}}A_{k_{i}}-\Gamma_{m\mu}^{n}k_{n}\partial_{k_{m}}A_{k_{i}}-\Gamma_{j\mu}^{i}A_{k_{j}}$\tabularnewline
\hline 
Bloch functions & $u(\boldsymbol{r},\boldsymbol{k};\boldsymbol{x},t)$ & $\nabla_{x^{\mu}}u\equiv\partial_{x^{\mu}}u-\Gamma_{m\mu}^{n}k_{n}\frac{\partial u}{\partial k_{m}}+\Gamma_{n\mu}^{m}r{}^{n}\frac{\partial u}{\partial r^{m}}$\tabularnewline
\hline 
\end{tabular}

\caption{\label{tab:lattice-covariant-derivative}lattice covariant derivative
of different phase space subjects}
\end{table*}

The mathematical expressions of the Berry connections involve derivatives
of the local Bloch states $u(\boldsymbol{r},\boldsymbol{\boldsymbol{k}};\boldsymbol{\boldsymbol{x}},t)$
with respect to the extended phase space parameters $(\boldsymbol{\boldsymbol{k}};\boldsymbol{\boldsymbol{x}},t)$.
However, this is non-trivial in the deforming crystal system. On top
of the special geometry of extended phase space mentioned before,
there is another difficulty due to Bloch states at different positions
and times have different periodicities in $\boldsymbol{r}$. A complete
understanding of this problem calls for the concept of Hilbert bundle
\cite{hilbertbundle}. Noticing that all the eigenstates of the local
Hamiltonian (\ref{7}) form a complete basis for the Hilbert space
of complex periodic functions with the same periodicity as the local
lattice, we can assign such a local Hilbert space to each position
$\boldsymbol{x}$, time $t$ and wave-vector $\boldsymbol{k}$. Then
we have a Hilbert bundle with its fibre the local Hilbert space denoted
by $\mathcal{F}(\boldsymbol{\boldsymbol{k}};\boldsymbol{x},t)$ and
the base manifold the extended phase space. $\mathcal{F}(\boldsymbol{\boldsymbol{k}};\boldsymbol{x},t)$
is characterized by the local periodicity given by $\{\boldsymbol{a}_{\alpha}(\boldsymbol{x},t)\}$.
Local Hamiltonian $\hat{H}(\boldsymbol{k};\boldsymbol{x},t)$ and
local Bloch states $u(\boldsymbol{r},\boldsymbol{k};\boldsymbol{x},t)$
are operator and states in $\mathcal{F}(\boldsymbol{k};\boldsymbol{x},t)$.
Thus the problem arises from comparing states in different Hilbert
spaces. 

Next, we discuss how to resolve this problem. Since Bloch functions
of all bands form a complete basis, for convenience we use them to
discuss the properties of states in the Hilbert bundle. Under a particular
choice of smooth gauge, the Bloch functions satisfy the following
boundary conditions:

\begin{align}
u(\boldsymbol{r}+\boldsymbol{a}_{\alpha}(\boldsymbol{x},t),\boldsymbol{k};\boldsymbol{x},t)= & u(\boldsymbol{r},\boldsymbol{k};\boldsymbol{x},t),\label{eq:34-1}\\
u(\boldsymbol{r},\boldsymbol{k}+2\pi\boldsymbol{b}^{\alpha}(\boldsymbol{x},t);x,t)= & e^{i\phi^{\alpha}(\boldsymbol{k};\boldsymbol{x},t)}e^{i2\pi\boldsymbol{r}\cdot\boldsymbol{b}^{\alpha}}u(\boldsymbol{r},\boldsymbol{k};\boldsymbol{x},t).\label{eq:35-1}
\end{align}
The first condition identifies the periodicity of the local Hilbert
space $\mathcal{F}(\boldsymbol{k};\boldsymbol{x},t)$, to which the
Bloch function belongs. The second condition shows that Bloch functions
are quasi-periodic functions in $\boldsymbol{k}$ where ``quasi''
is due to the Berry phase term $e^{i\phi^{\alpha}(\boldsymbol{k};\boldsymbol{x},t)}$.
The factor $e^{i2\pi\boldsymbol{r}\cdot\boldsymbol{b}^{\alpha}}$
is completely artificial because we denote the Brillouin zone torus
with a single domain $\boldsymbol{k}\in[0,2\pi b(\boldsymbol{x},t)]$.
The Berry phase term cannot be eliminated by single-valued gauge transformation.
For example, in two-dimensional case, the Berry phase accumulated
along the Brillouin zone boundary equals the Chern number. 

We view (\ref{eq:34-1}, \ref{eq:35-1}) as the boundary conditions
characterizing a Hilbert bundle state. Then it is natural to require
the correct derivative operation of quantum states is still a quantity
in this Hilbert bundle and satisfies the above boundary conditions.
It can be verified easily the gauge invariant derivative of $\boldsymbol{k}$
$\partial_{\boldsymbol{k}}+iA_{\boldsymbol{k}}$ satisfies this requirement,
where $A_{\boldsymbol{k}}$ is defined with $\partial_{\boldsymbol{k}}$
as $A_{\boldsymbol{k}}=\langle u|i\partial_{\boldsymbol{k}}u\rangle$.
However, the gauge invariant partial derivative of position $\boldsymbol{x}$
and time $t$ does not satisfy our requirement. Thus we introduce
the lattice covariant derivative in the Hilbert bundle denoted by
$\nabla_{x^{\mu}}$, whose property is given in Table \ref{tab:lattice-covariant-derivative}
and corresponding gauge invariant derivative $\nabla_{x^{\mu}}+iA_{x^{\mu}}$
satisfies the boundary conditions. Our discussion in last section
is the case where lattice covariant derivative acting on phase space
functions and is given in the first row. The last row in Table \ref{tab:lattice-covariant-derivative}
shows how lattice covariant derivative acting on Bloch functions.
Comparing the first row and the last row, we see that when lattice
covariant derivative acting on Bloch functions, in addition to the
first two terms which treat Bloch functions in the same way as phase
space functions, the third term resolves the issue the periodicity
difference of Bloch functions at different position and time in the
same manner. 

With lattice covariant derivative, the definition of Berry connection
is straight forward:
\begin{align}
A_{x^{\mu}}(\boldsymbol{k};\boldsymbol{x},t)\equiv & i\langle u(\boldsymbol{k};\boldsymbol{x},t)|\nabla_{x^{\mu}}u(\boldsymbol{k};\boldsymbol{x},t)\rangle,\label{eq:47}\\
A_{k}(\boldsymbol{k};\boldsymbol{x},t)\equiv & i\langle u(\boldsymbol{k};\boldsymbol{x},t)|\partial_{k}u(\boldsymbol{k};\boldsymbol{x},t)\rangle,\label{eq:48}
\end{align}
where the Bloch functions are normalized with the inner product:

\begin{align}
\langle u_{1}|u_{2}\rangle= & \frac{1}{v(\boldsymbol{x},t)}\int_{v}d^{3}ru_{1}^{*}(\boldsymbol{r},\boldsymbol{k};\boldsymbol{x},t)u_{2}(\boldsymbol{r},\boldsymbol{k};\boldsymbol{x},t),\label{eq:41}
\end{align}
where $u_{1}(\boldsymbol{r},\boldsymbol{k};\boldsymbol{x},t)$ and
$u_{2}(\boldsymbol{r},\boldsymbol{k};\boldsymbol{x},t)$ are two local
Bloch functions of different bands. Here because $\nabla_{x^{\mu}}u(\boldsymbol{r},\boldsymbol{k};\boldsymbol{x},t)$
is periodic in $\boldsymbol{r}$, the integral in a unit-cell becomes
reasonable. The factor $\frac{1}{v(\boldsymbol{x},t)}$ in Eq. (\ref{eq:41})
can be viewed as the volume measure and is position and time dependence,
which is another indication that lab frame is ``curved'' for electrons.
This factor is also necessary for lattice covariant derivative to
satisfy the Lebniz rule: $\nabla_{x^{\mu}}\langle u^{\prime}|u\rangle=\langle\nabla_{x^{\mu}}u^{\prime}|u\rangle+\langle u^{\prime}|\nabla_{x^{\mu}}u\rangle$,
where the total inner product on the left hand is treated as a phase
space function. 

The corresponding Berry curvature is defined with lattice covariant
derivative as
\begin{align}
\Omega_{k_{i}k_{j}}\equiv & i[\langle\partial_{k_{i}}u|\partial_{k_{j}}u\rangle-\langle\partial_{k_{j}}u|\partial_{k_{i}}u\rangle],\label{eq:53}\\
\Omega_{k_{i}x^{\mu}}\equiv & i[\langle\partial_{k_{i}}u|\nabla_{x^{\mu}}u\rangle-\langle\nabla_{x^{\mu}}u|\partial_{k_{i}}u\rangle],\label{eq:54-1}
\end{align}
$\Omega_{\boldsymbol{xx}}$ is a second order quantity, which will
not be discussed here. However, from the above definition the relation
between Berry curvatures $\Omega_{\boldsymbol{kx}}$ and Berry connections
is not so trivial. It turns out that the relation $\Omega_{k_{i}x^{\mu}}=\partial_{k_{i}}A_{x^{\mu}}-\nabla_{x^{\mu}}A_{k_{i}}$
is valid only if the lattice covariant derivative of Berry connection
$A_{k}$ is defined as
\begin{align}
\nabla_{x^{\mu}}A_{k_{i}}\equiv & \partial_{x^{\mu}}A_{k_{i}}-\Gamma_{m\mu}^{n}k_{n}\partial_{k_{m}}A_{k_{i}}-\Gamma_{j\mu}^{i}A_{k_{j}},\label{eq:54}
\end{align}
where the first two terms treat the Berry connection as a normal phase
space function. However, we have an additional term.If looking at
the form of Eq. (\ref{eq:32}) which is the covariant derivative of
a covector form $\boldsymbol{k}$, we can see that the last term in
Eq. (\ref{eq:54}) is a manifestation of the vector property of Berry
connection $A_{\boldsymbol{k}}$. We summarize this property in the
second row of Table \ref{tab:lattice-covariant-derivative}. It can
be easily checked that in deed $A_{\boldsymbol{k}}$ transforms in
the same way as the coefficient of a three-dimensional vector under
Newtonian coordinate transformation. Mathematically, this additional
term is due to the commutation relation $[\nabla_{x^{\mu}},\frac{\partial}{\partial k_{i}}]u=\Gamma_{m\mu}^{i}\frac{\partial}{\partial k_{cm}}u$,
where $u$ is the local Bloch function. It is necessary for the gauge
invariance of Berry curvatures. And Berry curvatures $\Omega_{k_{i}k_{j}}$,
$\Omega_{k_{i}x^{j}}$ can be viewed as second rank tensors. 

\subsection{Energy correction and complete equations of motion\label{subsec:Energy-correction-and}}

We have discussed the lattice covariant derivative of Bloch functions
and the corresponding Berry connections. A complete discussion should
also includes the property of lattice covariant derivative acting
on quantum operators such as the local Hamiltonian. This can be achieved
by imposing Leibniz rule such that

\begin{align}
\nabla_{x^{\mu}}[\hat{S}u](\boldsymbol{k};\boldsymbol{x},t)\equiv & (\nabla_{x^{\mu}}\hat{S})u(\boldsymbol{k};\boldsymbol{x},t)+\hat{S}(\boldsymbol{k};\boldsymbol{x},t)\nabla_{x^{\mu}}u,
\end{align}
where $\hat{S}(\boldsymbol{k};\boldsymbol{x},t)$ is some operator
in Hilbert bundle which keeps the periodicity of Bloch functions.
With the above requirement, we can directly define the deformation
potential operator $\hat{D}_{m}^{n}$ in arbitrary crystal system
as 

\begin{align}
\nabla_{x^{\mu}}\hat{H}(\boldsymbol{k};\boldsymbol{x},t)\equiv & \Gamma_{n\mu}^{m}(\boldsymbol{x},t)\hat{D}_{m}^{n}.
\end{align}
We find that for a generic lattice Hamiltonian as shown in Eq. (\ref{7}),
$\hat{D}_{m}^{n}=\mathcal{V}_{m}^{n}-\frac{1}{m_{e}}(\frac{1}{i}\frac{\partial}{\partial r^{m}}+k_{m})(\frac{1}{i}\frac{\partial}{\partial r^{n}}+k_{n})$,
which is exactly the one appearing in Eq. (\ref{eq:19-1}). Historically,
deformation potential operator is first derived using Lagrangian coordinate
\cite{PhysRev.121.720}. Here we show that it is simply the lattice
covariant derivative of local Hamiltonian. This conclusion only relies
on the existence of one to one correspondence between ionic distribution
and local Bravais lattice. 

With lattice covariant derivative, most well known results have the
same analytical form only with partial derivatives replace by lattice
covariant derivatives. For example the Hellmann\textendash Feynman
theorem in the deformation crystal case can be written as:
\begin{align}
\nabla_{x^{\mu}}\varepsilon(\boldsymbol{k};\boldsymbol{x},t) & =\langle u(\boldsymbol{k};\boldsymbol{x},t)|\hat{D}_{n}^{m}|u(\boldsymbol{k};\boldsymbol{x},t)\rangle\Gamma_{m\mu}^{n},\label{eq:45}
\end{align}
where the left hand side is to treat the local band energy as a phase
space function and the right hand side comes from the lattice covariant
derivative of the expectation values of local Hamiltonian in Bloch
states. Comparing to Eq. (\ref{eq:28}), it is obvious that deformation
potential $D_{n}^{m}$ is just the expectation value of the deformation
potential operator.

Next we discuss the first order correction to energy. It contains
static part and dynamical part. The static contribution comes from
the expectation value of the gradient correction Eq. (\ref{eq:19-1})
in the wave-packet states, which has two terms: the potential correction
$\Gamma_{jk}^{i}(\boldsymbol{x},t)\langle u|\hat{O}_{i}^{jk}|u\rangle$
and the gradient correction $\Gamma_{ni}^{m}\boldsymbol{Im}[\langle u|D_{mn}-\hat{D}_{mn}|\partial_{k_{i}}u\rangle]$.
The dynamical part comes from the coupling between lattice rotation
and self-rotation of wave-packet: $2\boldsymbol{\omega\cdot J},$
which is a Zeeman type coupling with $\boldsymbol{J}=m_{e}\frac{i}{2\hbar}\langle\partial_{\boldsymbol{k}}u|\times(\varepsilon-\hat{H})|\partial_{\boldsymbol{k}}u\rangle$
the angular momentum of wave-packet. This can be understood from the
similarity between the form of Coriolis force in Eq. (\ref{eq:31})
to Lorentz force. The detailed derivation of these correction terms
is given in Appendix \ref{sec:Lagrangian-of-wave-packet}. In summary,
the total energy up to first order reads:
\begin{align}
\varepsilon_{tot}=\varepsilon+\boldsymbol{\Gamma}_{i}\operatorname{Im}\langle u|\boldsymbol{D}-\hat{\boldsymbol{D}}|\partial_{k_{i}}u\rangle+\Gamma\langle\hat{O}\rangle+2\boldsymbol{\omega}\cdot\boldsymbol{J}.
\end{align}

Up till now, we have all the ingredients to write down the equations
of motion up to first order. They are achieved by adding Berry curvatures
to the Eq. (\ref{eq:30}), (\ref{eq:31}) and using the total energy
instead of local band energy. The results are
\begin{align}
D_{t}\boldsymbol{x}= & \partial_{\boldsymbol{k}}\varepsilon_{tot}-D_{t}\boldsymbol{k}\times\Omega_{\boldsymbol{k}}-\Omega_{\boldsymbol{k}T},\label{eq:51}\\
D_{t}\boldsymbol{k}= & -\nabla_{\boldsymbol{x}}\varepsilon_{tot}+m_{e}D_{t}\boldsymbol{x}\times2\boldsymbol{\omega}-m_{e}\boldsymbol{a},\label{eq:52}
\end{align}
which have the similar analytical form as the result in \cite{sundaram}.
However, partial derivatives are replace by lattice covariant derivative
and the geometrical meaning of each term is much clearer. Eq. (\ref{eq:51})
is the relative velocity of electron to ions and each term can be
viewed as a spatial vector while Eq. (\ref{eq:52}) is the mechanical
change of crystal momentum and each term can be viewed as a spatial
covector. This geometrical property guarantees the covariant form
of the equations of motion and allows us to directly write down the
equations of motion in other coordinates related by Newtonian coordinate
transformation. $\Omega_{\boldsymbol{k}}$ is the pseudo vector constructed
from $\Omega_{\boldsymbol{kk}}$. $\Omega_{\boldsymbol{k}T}=\Omega_{\boldsymbol{kx}}\cdot\dot{\boldsymbol{x}}+\Omega_{\boldsymbol{k}t}$
will gives rise to the adiabatic current induced by strain rate and
strain gradient. And the above equations of motion can be derived
from the complete first order Lagrangian:
\begin{align}
L_{e}= & -(\varepsilon_{tot}+\boldsymbol{W}\cdot\boldsymbol{k}+\frac{1}{2}m_{e}\boldsymbol{W}^{2})+(\boldsymbol{k}+m_{e}\boldsymbol{W})\cdot\dot{\boldsymbol{x}}\nonumber \\
 & +(A_{t}+\dot{\boldsymbol{x}}\cdot A_{\boldsymbol{x}})+D_{t}\boldsymbol{k}\cdot A_{\boldsymbol{k}}.\label{eq:83-1}
\end{align}

\subsection{Applications\label{subsec:Applications}}

\begin{table*}[!t]
\begin{tabular}{|c|>{\centering}p{7cm}|>{\centering}p{7cm}|}
\hline 
 & examples & lattice covariant derivatives\tabularnewline
\hline 
phase space functions & $\varepsilon(\boldsymbol{k};\boldsymbol{x},t)$ & $\nabla_{n}^{m}\varepsilon\equiv a_{\alpha}^{m}\frac{\partial\varepsilon}{\partial a_{\alpha}^{n}}-\Gamma_{m\mu}^{n}k_{n}\frac{\partial\varepsilon}{\partial k_{m}}$\tabularnewline
\hline 
phase space vectors  & $A_{k}(\boldsymbol{k};\boldsymbol{x},t)$ & $\nabla_{n}^{m}A_{k_{i}}\equiv a_{\alpha}^{m}\frac{\partial A_{k_{i}}}{\partial a_{\alpha}^{n}}-k_{n}\partial_{k_{m}}A_{k_{i}}-\delta_{n}^{i}A_{k_{m}}$\tabularnewline
\hline 
Bloch functions & $u(\boldsymbol{r},\boldsymbol{k};\boldsymbol{x},t)$ & $\nabla_{n}^{m}u\equiv a_{\alpha}^{m}\frac{\partial u}{\partial a_{\alpha}^{n}}-k_{n}\frac{\partial u}{\partial k_{m}}+r{}^{n}\frac{\partial u}{\partial r^{m}}$\tabularnewline
\hline 
\end{tabular}

\caption{\label{tab:covariant-strain-derivative}covariant strain derivative
of different phase space subjects}
\end{table*}

In this section, we discuss the piezoelectricity, flexoelectricity,
strain rate induced orbital magnetization and electron stress tensor
as well as their responses to deformation. Frequently, we need to
extract a factor which is related to deformation e.g. strain, strain
gradient and velocity gradient to get the corresponding response coefficient.
This is achieved by defining the covariant strain derivative $\nabla_{n}^{m}$,
which has the physical meaning as the differentiation to the unsymmetrized
strain tensor and is related to lattice covariant derivative as $\nabla_{x^{\mu}}=\Gamma_{n\mu}^{m}\nabla_{n}^{m}$.
However, we should notice that this is true only when the position
and time dependence of the system comes from the lattice vectors $\{\boldsymbol{a}_{\alpha}(\boldsymbol{x},t)\}$
such that $\partial_{x^{\mu}}=\Gamma_{n\mu}^{m}a_{\alpha}^{n}\partial_{a_{\alpha}^{m}}$
in Table \ref{tab:lattice-covariant-derivative}. Similarly, we summarize
its action on various phase space quantities in Table \ref{tab:covariant-strain-derivative}. 

\subsubsection{deformation induced adiabatic charge current}

For simplicity, in the discussion of deformation induced adiabatic
charge current, we only consider band insulators at zero temperature
such that the distribution function is simply a step function. First
we study the total current of both ions and electrons under clamped-ion
approximation. This approximation states that at each time the distribution
of ions within a unit-cell follows the instantaneous strain of the
unit-cell (although there can be internal strain contribution \cite{Tagantsevflexo}).
The first point to notice that in general a deformed band insulator
is not locally charge neutral due to inhomogeneous electric polarization.
Particularly, the electron charge density is given by the integration
of density of state $-e\sum\int d^{3}k[1+tr(\Omega_{\boldsymbol{kx}})-m_{e}2\boldsymbol{\omega}\cdot\Omega_{\boldsymbol{k}}]$
\cite{DOS}, where the summation is over all occupied bands. Under
periodic gauge of $A_{\boldsymbol{x}}$ in $\boldsymbol{k}$, it reads
$-\frac{en_{e}}{v}-\partial_{i}P^{i}+em_{e}2\boldsymbol{\omega}\cdot\boldsymbol{\Omega}$,
where $\boldsymbol{\Omega}=\frac{1}{(2\pi)^{3}}\int d^{3}k\Omega_{\boldsymbol{k}}$
is quantized and $\boldsymbol{P}=-\frac{e}{(2\pi)^{3}}\sum\int d^{3}kA_{\boldsymbol{k}}$
is the Vanderbilt polarization \cite{polarization} and $n_{e}$ the
number of itinerate electrons per unit-cell. The first term is cancelled
by the ionic charge density. Next we show that for band insulators
at zero temperature, the total current is adiabatic and can be categorized
as either electric polarization current or electric magnetization
current besides the anomalous current. Particularly, the magnetization
due to the motion of polarization dipole modifies charge pumping picture
and leads to the concept of proper piezoelectricity \cite{pieozoelectricity,Vanderbiltproperpolarization}.

Using the equations of motion, it is straight forward to write down
the total current density up to first order (Appendix \ref{sec:Orbital-magnetization-and}):

\begin{align}
\boldsymbol{j}_{tot}= & -em_{e}\boldsymbol{a}\times\boldsymbol{\Omega}+\partial\times\boldsymbol{M}+\partial_{t}\boldsymbol{P}+\partial\times(\boldsymbol{P}\times\boldsymbol{W})\label{eq:49-1}
\end{align}
where $e$ is the absolute value of electron charge and the integral
is in local $\boldsymbol{k}$ space. And the periodic gauge for the
real space Berry connections: $A_{x^{\mu}}(\boldsymbol{k})=A_{x^{\mu}}(\boldsymbol{k}+2\pi\boldsymbol{b})$
is used. The first term is the well known anomalous current density
\cite{karplus1954hall}, with $\boldsymbol{\Omega}=\frac{1}{(2\pi)^{3}}\int d^{3}k\Omega_{\boldsymbol{k}}$.
However, the anomalous current density is driven by the inertial force
due to ionic acceleration. The second term is the magnetization current
density, where $\boldsymbol{M}=-\frac{1}{(2\pi)^{3}}\frac{e}{2\hbar}\sum_{n}\int i\langle\partial_{\boldsymbol{k}}u_{n}|\times(\varepsilon_{n}+\hat{H}-2\mu)|\partial_{\boldsymbol{k}}u_{n}\rangle$
is the orbital magnetization at zero temperature \cite{orbitalmagnetization}.
The third term is the polarization current density. Attention should
be paid to the last term. It is the curl of $\boldsymbol{P}\times\boldsymbol{W}$,
which indicates that $\boldsymbol{P}\times\boldsymbol{W}$ should
be interpreted as the magnetization density. This term comes from
the motion of a polarized material, which is actually a well known
phenomenon in classical electro-magnetism. As magnetization and polarization
form an 3+1 dimensional anti-symmetric tensor, they transform into
each other under material motion. The picture is as follows: consider
an initially stationary dipole moment $\boldsymbol{P}$ composed by
a pair of spatially separated positive and negative charge, if it
begins to move at velocity $\boldsymbol{W}$, the two charges give
rise to currents of opposite directions thus an effective current
circuit is formed which gives rise to the orbital magnetization density
$\boldsymbol{P}\times\boldsymbol{W}$. This effect shows the consistency
between the classical electro-magnetism and the modern quantum theory
of polarization and magnetization in solids. 

We then consider the case where the lattice vectors are constant in
space and only changes slowly with time and further assume that at
each fixed time $t$, the lattice has time-reversal symmetry. Then
the first term and second term in Eq. (\ref{eq:49-1}) vanish and
using the elastic condition (\ref{eq:12-1}, \ref{eq:13-1}) the remaining
terms can be written in an intriguing form as

\begin{align}
j_{tot}^{i}= & \partial_{t}P^{i}-\Gamma_{m0}^{i}P^{m}+\Gamma_{m0}^{m}P^{i}.\label{eq:52-1}
\end{align}
The first term is the absolute change of polarization density under
deformation. The second term is the directional change due to the
deformation of crystalline directions. The last term is the magnitude
change due to the deformation of unit-cell volume. Eq. (\ref{eq:52-1})
shows that the last two geometrical changes of polarization density
should be subtracted to give the experimentally observed current density.
This confirms the argument by Nelson and Vanderbilt \cite{Nelsonproperpieozoelectricity,Vanderbiltproperpolarization}
that only the proper change of polarization can be observed experimentally.
Substituting $\boldsymbol{P}=-\frac{e}{(2\pi)^{3}}\int d^{3}kA_{\boldsymbol{k}}$
and $\partial_{t}=\Gamma_{n0}^{m}a_{\alpha}^{n}\partial_{a_{\alpha}^{m}}$,
within our lattice covariant formula, Eq. (\ref{eq:52-1}) can be
conveniently written as

\begin{align}
j^{i}= & -e\Gamma_{m0}^{n}\int\frac{d^{3}k}{(2\pi)^{3}}[\nabla_{n}^{m}A_{k_{i}}-\partial_{k_{i}}A_{n}^{m}],\label{eq:52-1-1}
\end{align}
where the periodic gauge condition for $A_{t}=\Gamma_{m0}^{n}A_{n}^{m}$
is used again to retain this gauge invariant form and $A_{n}^{m}=i\langle u|\nabla_{n}^{m}u\rangle$.
Noticing $\boldsymbol{\Gamma}_{0}$ denotes the strain rate and by
defining the proper piezoelectric constant as $e_{n}^{mi}=\frac{\partial j^{i}}{\partial\Gamma_{m0}^{n}}$,
we have an explicit expression for $e_{n}^{mi}$ as 

\begin{align}
e_{n}^{mi}= & -e\int\frac{d^{3}k}{(2\pi)^{3}}[i\langle\partial_{k_{i}}u|\nabla_{n}^{m}u\rangle-i\langle\nabla_{n}^{m}u|\partial_{k_{i}}u\rangle],
\end{align}
which is nothing but the integral of Berry curvature involving $\boldsymbol{k}$
and strain. And lattice covariant strain derivative gives an explicit
meaning for the strain derivative. Its expression in terms of deformation
potential operator is discussed in Appendix \ref{sec:Expressions-in-momentum}. 

Another consequence of Eq. (\ref{eq:52-1}) is that the charge pumping
picture should be revised in deforming crystal case. The usual picture
states that when the system varies slowly and periodically in time,
the charge pumped through a fixed plane in lab frame during one cycle
is quantized \cite{chargepumping}. However, due to the last two terms
in Eq. (\ref{eq:52-1}), this picture is changed due to that the right
hand side is not a total time derivative. To see this, by multiplying
a factor $v\boldsymbol{b}^{\alpha}$ on both sides, Eq. (\ref{eq:52-1})
reads:
\begin{align}
j_{tot}^{i}vb_{i}^{\alpha}= & -\partial_{t}(vb_{i}^{\alpha}P^{i}),\label{eq:59}
\end{align}
where $v(t)$ is the unit-cell volume and $\boldsymbol{b}^{\alpha}(t)$
is the reciprocal lattice vectors. Now the right hand side becomes
a total time derivative and the left hand side is the current passing
through a lattice plane unit-cell, where $\boldsymbol{b}^{\alpha}$
is the normal direction. After integration in time, the left hand
side gives the total charge pumped through a lattice plane unit-cell
and the right hand side is the difference of $vb_{l}^{\alpha}P^{l}$
between initial and final state. Suppose the initial and final states
are the same, from the uncertainty of $P^{l}$, the charge pumped
is quantized to some integer. However, noticing that the lattice plane
unit-cell is constantly changing during the time cycle, the pumped
charge through a fixed surface plane in lab frame is not necessarily
quantized.

\subsubsection{Strain gradient induced polarization and strain rate induced magnetization}

\begin{table*}[!t]
\begin{tabular}{|>{\centering}p{5cm}|>{\centering}m{3cm}|c|}
\hline 
 & polarization  & \multicolumn{1}{>{\centering}p{3cm}|}{orbital magnetization}\tabularnewline
\hline 
intrinsic contribution & $\boldsymbol{P}_{in}=e\int A_{\boldsymbol{k}}$ & $\boldsymbol{M}_{in}=\boldsymbol{M}+\boldsymbol{P}\times\boldsymbol{W}$\tabularnewline
\hline 
Chern-Simons contribution & $\boldsymbol{P}_{cs}^{i}=e\Gamma_{mj}^{n}\mu_{n}^{mij}$ & $M_{cs}^{ij}=-e\partial_{m}W^{n}\mu_{n}^{mij}+(P_{cs}^{i}W^{j}-P_{cs}^{j}W^{i})$\tabularnewline
\hline 
\end{tabular}

\caption{\label{tab:electric-polarization-and}electric polarization and orbital
magnetization}
\end{table*}

The first order current density comes from the variation of the zeroth
order polarization/magnetization. To study the polarization/magnetization
induced by strain gradient/rate, current density accurate to second
order is needed. As pointed in the work \cite{inhomogeneouspolarization},
polarization/magnetization due to inhomogeneity can be divided into
two parts: (1) the zeroth order polarization/magnetization formula
expressed with inhomogeneity modified Bloch functions; (2) the Chern-Simons
contribution expressed with zeroth order local Bloch function. The
former will be deferred to future study. Here we concentrate on the
Chern-Simons contribution to polarization/magnetization from electrons. 

Two results are discussed here: the polarization induced by the strain
gradient denoted by lattice connection and orbital magnetization induced
by strain rate denoted by the gradient of velocity field. The former
phenomenon is well known as flexoelectricity\cite{flexoelectrictensor}
while the latter phenomenon we call dynamical magnetization. We find
that the Chern-Simons contribution to both effects share the same
response tensor coefficient given by 

\begin{align}
\mu_{n}^{mij}= & \int[A_{k_{i}}\nabla_{n}^{m}A_{k_{j}}+A_{k_{j}}\partial_{k_{i}}A_{n}^{m}+A_{n}^{m}\partial_{k_{j}}A_{k_{i}}],\label{eq:53-2}
\end{align}
which is just the $\boldsymbol{k}$ space integral of a Chern-Simons
form involving one strain parameter and two $\boldsymbol{k}$ parameter.
$A_{n}^{m}$ is the Berry connection in terms of strain.

The tensor coefficient $\mu_{n}^{mij}$ is anti-symmetric respect
to the indices $ij$ and symmetric to $mn$. The former property is
inherited from Chern-Simons form and the latter is due to the vanishing
of lattice covariant derivative of Bloch functions under local rotation.
In terms of this coefficient, the Chern-Simons polarization and magnetization
induced by strain gradient and strain rate respectively can be written
as

\begin{align}
P_{cs}^{i}= & e\Gamma_{mj}^{n}\mu_{n}^{mij},\label{eq:53-1}\\
M_{cs}^{ij}= & -e\partial_{m}W^{n}\mu_{n}^{mij}+(P_{cs}^{i}W^{j}-P_{cs}^{j}W^{i}).\label{eq:54-3}
\end{align}
Eq. (\ref{eq:53-1}) is the Chern-Simons contribution to flexoelectricity,
where the strain gradient is denoted by lattice connection $\{\Gamma_{mj}^{n}\}$.
The well known picture to understand flexoelectricity is introduced
by Taganstsev, where flexoelectricity is described by the ionic effective
Born charge multiplied by the displacement induced by strain gradient
\cite{Tagantsevflexo}. The major challenge is the calculation of
the effective charge. The longitudinal polarization can be calculated
from the local charge density response to ionic position \cite{pieozoelectricity,jhong}.
However, the transverse part involves the current response to strain
gradient\cite{jhong}. As can be seen from Appendix \ref{sec:Orbital-magnetization-and},
in deed we achieve the above formula by considering the response of
current to strain gradient and strain rate. The key point is to calculate
the Abelian Chern-Simons form (\ref{eq:53-2}), which is a tractable
problem. 

Eq. (\ref{eq:54-3}) is the dynamical magnetization as clearly indicated
by the appearance of velocity field. The first term is induced by
the gradient of velocity field, which is the strain rate experienced
by ions. It couples to the same tensor coefficient of Chern-Simons
flexoelectricity. This indicates that materials with large bulk flexoelectricity
effect may also demonstrate observable dynamical magnetization. The
second term is the transformation from polarization to magnetization
due to ionic motion as discussed before. 

We conclude by summarizing the different parts of polarization/magnetization
in Table \ref{tab:electric-polarization-and}, which includes the
zeroth order contributions intrinsic to local lattices and the Chern-Simons
contribution due to inhomogeneity. 

\subsubsection{Stress tensor and its responses }

The electron stress tensor response to geometrical background is a
very interesting problem. Particularly, the response to velocity gradient
is known as viscosity term and is a manifestation of the rigidity
of the electron system. Electron viscosity has been studied in different
cases such as the integer Hall system \cite{hallviscosity,hallviscosity1},
fractional quantum Hall \cite{FQHviscosity,Vignale,Vignaletokatly},
topological insulators \cite{viscosityTI}, superfluid\cite{viscositysuperfluid}
and in the time dependent DFT theory \cite{VignaleDFTViscosity}.
Here we give a general formula of electron energy stress tensor in
a spatially homogeneous band insulator at zero temperature. Its response
to lattice deformation is explicitly shown in the following form:

\begin{widetext} 

\begin{align}
\mathcal{T}_{j}^{i}= & \mathcal{D}_{j}^{i}+2\omega\cdot\nabla_{j}^{i}\mathcal{J}+\Gamma_{m0}^{n}\eta_{nj}^{mi}+a\cdot\nabla_{j}^{i}P_{m_{e}}+\frac{m_{e}}{v}W{}^{i}W{}^{j}-[W^{j}j_{m_{e}}^{i}+i\leftrightarrow j].\label{eq:58}
\end{align}

\end{widetext} where the left hand side is the stress energy tensor.
The derivation is given in Appendix \ref{sec:Energy-stress-tensor}.
First we would like to point out that the indices $i$ and $j$ are
symmetric on both sides of the equation. This is because energy stress
tensor can be viewed as the unsymmetrized strain derivative of electron
energy. And this covariant strain derivative vanishes when strain
is antisymmetric i.e crystal under rigid body rotation. 

The above expression is for a particular filled band while the total
energy stress tensor is the sum of all occupied bands. $\mathcal{D}_{j}^{i}=\int\frac{d^{3}k}{(2\pi)^{3}}D_{j}^{i}$
is the contribution from deformation potential. It gives the leading
order contribution to stress tensor. The second term is the response
of stress tensor to the rotation of lattices in time, which is the
anti-symmetric part of ionic velocity gradient. $\mathcal{J}$ is
the angular momentum for a filled band. 

\begin{align}
 & \nabla_{i}^{j}\mathcal{J}^{mn}=a_{\alpha}^{j}\frac{\partial\mathcal{J}^{mn}}{\partial a_{\alpha}^{i}}-\delta_{i}^{m}\mathcal{J}^{jn}-\delta_{i}^{n}\mathcal{J}^{mj}+\delta_{i}^{j}\mathcal{J}^{mn},\label{eq: 61}
\end{align}
where $\mathcal{J}^{mn}=m_{e}\int\frac{d^{3}\boldsymbol{k}}{(2\pi)^{3}}\operatorname{Im}\langle\partial_{k_{m}}u\mid(\varepsilon+\hat{H})\mid\partial_{k_{n}}u\rangle$.
The first three terms in Eq. (\ref{eq:61}) are the strain derivative
of a second rank tensor in retrospect to the first rank vector case
exemplified by Berry connection term in Table \ref{tab:covariant-strain-derivative}.
The last term comes from the strain derivative of the volume element
in $\boldsymbol{\boldsymbol{k}}$ space, which reads $\nabla_{n}^{m}(d^{3}k)=-\delta_{n}^{m}d^{3}\boldsymbol{k}$.
If we are to put strain derivative outside of the integral $\int d^{3}\boldsymbol{k}$,
this term always appears. Equivalently, we can view a phase space
quantity after integration in $\boldsymbol{k}$ space as a real space
density quantity. The last term is a manifestation of this density
property. 

The fourth term in Eq. (\ref{eq:58}) is the response to strain rate,
which is often referred to as the viscosity term. The viscosity tensor
has the following explicit form as:

\begin{align}
\eta_{nj}^{mi}=\int & d^{3}k[i\langle\nabla_{n}^{m}u|\nabla_{j}^{i}u\rangle-i\langle\nabla_{j}^{i}u|\nabla_{n}^{m}u\rangle],
\end{align}
where is simply the integral of Berry curvature in terms of strain
parameter in $\boldsymbol{k}$ space. Again the meaning of strain
derivative is only clear within our theory as given in Table \ref{tab:covariant-strain-derivative}.
This term is automatically antisymmetric between the two groups of
indices $mn$ and $ij$, thus is dissipationless. And both $mn$ and
$ij$ are symmetric within their own groups. This symmetric property
is inherited from the fact that $|\nabla_{n}^{m}u\rangle=0$ when
$m,n$ are anti-symmetric i.e rigid body rotation. Its expression
in terms of deformation potential operator is given in Appendix \ref{sec:Expressions-in-momentum}.

The third term in Eq. (\ref{eq:58}) is the response to acceleration.
$P_{m_{e}}=m_{e}\int d^{3}kA_{k}$ is the mass polarization of electron
and

\begin{align}
\nabla_{i}^{j}P_{m_{e}}^{n}= & a_{\alpha}^{j}\frac{\partial P_{m_{e}}^{n}}{\partial a_{\alpha}^{i}}-\delta_{i}^{n}P_{m_{e}}^{j}+\delta_{i}^{j}P_{m_{e}}^{n}.
\end{align}
Comparing to Eq. (\ref{eq: 61}), it is easily to see this is the
strain derivative of a vector density. Noticing the periodic gauge
condition for $A_{n}^{m}$, actually $\nabla_{i}^{j}P_{m_{e}}^{n}=\frac{m_{e}}{e}e_{j}^{in}$
is just the proper piezoelectric constant multiplied by a factor $\frac{m_{e}}{e}$. 

The last two terms Eq. (\ref{eq:58}) has the signature of a perfect
fluid with energy density and energy current density $(\frac{m_{e}}{v},-j_{m_{e}})$
in coordinates relative to ions. After transforming back to lab frame
with relative velocity $-\boldsymbol{W}$ , we have those terms as
part of the energy stress tensor. And $\boldsymbol{j}_{m_{e}}$ is
the mass current density with expression $j_{m_{e}}^{i}=\frac{m_{e}}{e}\Gamma_{n0}^{m}e_{m}^{ni}+\frac{m_{e}W^{i}}{v}-m_{e}^{2}(\boldsymbol{a}\times\boldsymbol{\Omega})^{i}+(\boldsymbol{\partial}\times\boldsymbol{\mathcal{J}})^{i}$
in the spatial homogeneous case. 

Here we only discussed the spatial homogeneous case for band insulators.
The physical meaning of the above stress energy tensor is to provide
a force effect on the dynamics of ions as shown in the derivation
Appendix \ref{sec:Energy-stress-tensor}. It is also interesting to
study the hydrodynamics of electrons in a metal under deformation
to see the effect of energy stress tensor on the equations of motion.
And in that case we need to solve the Botlzman equation in combination
with the equations of motion Eq. (\ref{eq:51}, \ref{eq:52}). And
we didn't include the strain gradient contribution to the energy stress
tensor, which will be referred to our next paper. 

\section{CONCLUSION}

In summary, we have developed a theory describing the semiclassical
dynamics of electrons in deforming crystals up to the first order
of strain gradient, strain rate and lattice acceleration. Our theory
is based on lattice bundle picture, where local lattices are introduced
to account for the local property of deforming crystals. To compare
quantities associated with local lattices with different periodicities,
a derivative operation called lattice covariant derivative is introduced.
It takes the place of partial derivative in expressing the equations
of motion including Berry phase effect. In general, lattice covariant
derivative allows our results expressed in a familiar and covariant
form under Newtonian coordinate transformation. The geometrical effect
of lattice deformation is made explicit in terms of our lattice covariant
formalism. Many deformation effects resemble the effects in a curved
spacetime even if expressed in the Euclidean lab frame coordinates. 

Our formula considers the deformation of an original periodic Hamiltonian
and makes no other particular assumption about the property of the
Hamiltonian. Thus we expect the results can be easily applied to other
periodic systems such as the photonic crystal or cold atom systems.
Moreover, our approach provides a way to generate non-trivial geometry
for particles coupled to a deformed background. In other systems of
different order parameter (in our case the lattice vectors), we expect
other types of geometry can be achieved. As the focus of this paper
is to set up the frame work of our lattice covariant formula, a lot
of discussions in the application part is far from complete. The most
obvious direction to pursue is to include the strain gradient contribution
to Bloch functions and the electron energy stress tensor. 
\begin{acknowledgments}
The authors thank Junren Shi, Ji Feng, Michael Stone for insightful
discussions, Cong Xiao and Xiao Li for revising this paper. This work
was supported by China 973 Program (2013CB921900 and 2012CB921300).
\end{acknowledgments}

\appendix

\section{Lattice frame\label{sec:Lattice-frame}}

In the main text, the results are expressed in lab frame. This has
the advantage that the physical picture is more transparent. However,
it is more convenient to derive the results in another curvilinear
frame called lattice frame. This is similar to the relation between
Euclidean and Lagrangian description in fluid dynamics. This coordinate
transformation method in dealing with deforming crystal problem is
introduced by Whitfield \cite{PhysRev.121.720}.

Here we introduce the lattice frame and discuss its relation with
the lattice bundle picture. Lattice frame coordinates are denoted
as $\{x^{\prime\alpha},t^{\prime}\}$, with $\alpha=1,2,3$ representing
three crystalline directions. Given the positions of all lattice points
in lab frame $\{\boldsymbol{R}_{\boldsymbol{l}}(t)\}$, we define
a smooth lattice field $\boldsymbol{R}(\boldsymbol{x}^{\prime},t^{\prime})$
in terms of lattice frame coordinates, which satisfies:

\begin{align}
\boldsymbol{R}_{\boldsymbol{l}}(t)\equiv & \boldsymbol{R}(\boldsymbol{l},t^{\prime})|_{t=t^{\prime}}.\label{eq:1}
\end{align}
 Then lab frame coordinates is related to lattice frame coordinates
as 

\begin{align}
\boldsymbol{x}= & \boldsymbol{R}(\boldsymbol{x}^{\prime},t^{\prime}),\label{eq:2}\\
t= & t^{\prime}.
\end{align}
From (\ref{eq:1}), we see that lattice frame coordinates can be viewed
as the continuity of the lattice points label. Any deforming crystal
is mapped to a unit cubic lattice in lattice frame. We require that
in a local region deformation is elastic which means the relation
(\ref{eq:2}) is reversible and $\boldsymbol{x}^{\prime}$ is also
a function of $\boldsymbol{x}$ and $t$. In later discussion , we
will frequently use this reversibility and change the independent
variables of the same fields from $(\boldsymbol{x}^{\prime},t^{\prime})$
to $(\boldsymbol{x},t)$ or vice versa.

To connect to the lattice bundle picture introduced in the main body
of this paper, we can define the lattice vector fields and velocity
field as
\begin{align}
\boldsymbol{a}_{\alpha}(\boldsymbol{x}^{\prime},t^{\prime})\equiv & \partial_{x^{\prime\alpha}}\boldsymbol{R}(\boldsymbol{x}^{\prime},t^{\prime}),\label{eq:3-1}\\
\boldsymbol{W}(\boldsymbol{x}^{\prime},t^{\prime})\equiv & \partial_{t^{\prime}}\boldsymbol{R}(\boldsymbol{x}^{\prime},t^{\prime}).\label{eq:63}
\end{align}
The physical property of lattice vector fields and velocity field
comes naturally from (\ref{eq:1}) that

\begin{align}
\boldsymbol{a}_{\alpha}(\frac{\boldsymbol{l}+(\boldsymbol{l}+1^{\alpha})}{2},t^{\prime})= & \boldsymbol{R}_{\boldsymbol{l}+1^{\alpha}}(t^{\prime})-\boldsymbol{R}_{\boldsymbol{l}}(t^{\prime}),\label{eq:64}\\
\boldsymbol{W}(\boldsymbol{l},t^{\prime})= & \dot{\boldsymbol{R}}_{\boldsymbol{l}}(t^{\prime}).
\end{align}

However, from the definition (\ref{eq:1}), Eq. (\ref{eq:64}) does
not necessarily holds exactly. Here we impose the second requirement
for lattice field such that the LHS and RHS of Eq. (\ref{eq:64})
equal. This is to keep our theory accurate in the first order gradient
of lattice fields at least in the case of constant strain gradient.
In the constant strain gradient case, we have the following form of
lattice field $\boldsymbol{R}(\boldsymbol{x}^{\prime},t^{\prime})$: 

\begin{align}
\boldsymbol{R}(\boldsymbol{x}^{\prime},t^{\prime})= & \boldsymbol{C}_{\alpha}^{1}(t^{\prime})x^{\prime\alpha}+\frac{1}{2}\boldsymbol{C}_{\alpha\beta}^{2}(t^{\prime})x^{\prime\alpha}x^{\prime\beta}+\boldsymbol{C}^{0}(t^{\prime}),\label{eq: 68}
\end{align}
where $\boldsymbol{C}_{\alpha\beta}^{2}$ describes a constant strain
gradient in space. After substituting Eq. (\ref{eq: 68}) into Eq.
(\ref{eq:64}), we can see that Eq. (\ref{eq:64}) holds exactly. 

From the above definition of lattice vector fields and velocity fields,
it is straightforward to show that

\begin{align}
(\boldsymbol{a}_{\alpha}\cdot\boldsymbol{\partial})\boldsymbol{a}_{\beta}-(\boldsymbol{a}_{\beta}\cdot\boldsymbol{\partial})\boldsymbol{a}_{\alpha}= & 0,\label{5-1}\\
(\boldsymbol{a}_{\alpha}\cdot\boldsymbol{\partial})\boldsymbol{W}-(\boldsymbol{W}\cdot\boldsymbol{\partial})\boldsymbol{a}_{\alpha}= & \partial_{t}\boldsymbol{a}_{\alpha},
\end{align}
which is the elastic relation addressed by Eq. (\ref{5}\ref{6}).
This shows the consistency between the definition here and the discussion
in the main body of this paper. Actually elastic condition is the
necessary condition for the existence of a local lattice fields $\boldsymbol{R}(x^{\prime},t)$. 

Next we discuss the metrics in lattice frame. In lattice frame, lattice
points always have unit cubic lattice coordinates and the deformation
is described by the metric tensor in this curvilinear coordinate system.
This is contrast to lab frame description where coordinate of lattice
points are crucial. The four-dimensional metric tensor in lattice
frame is

\begin{align}
G_{\alpha\beta}= & \frac{\partial x^{i}}{\partial x^{\prime\alpha}}\frac{\partial x^{i}}{\partial x^{\prime\beta}}=a_{\alpha}^{i}a_{\beta}^{i},\label{eq:68-1}\\
G_{0\alpha}= & \frac{\partial x^{i}}{\partial t^{\prime}}\frac{\partial x^{i}}{\partial x^{\prime\alpha}}=W^{i}a_{\alpha}^{i},\label{eq:69-1}
\end{align}
where we choose $(-1,1,1,1)$ for the Minkovski metric signature.
We see that from the above expression the spatial part of metric is
just the contraction between two lattice vectors and the time-space
component of metric is just the velocity field projected to lattice
frame coordinate directions. This is consistent with the geometric
method of describing deformation effect \cite{threedimensioanlgravity,Bilby263}
where metric tensor is introduced to account for strain effect. Next,
we discuss how the metric fields couple to the first principle Hamiltonian
of electrons written in lattice frame.

\section{Gradient expansion to crystal potential \label{sec:Gradient-expansion-to}}

Viewed in lab frame, the total crystal potential which depends on
deformed lattice points is responsible for all the deformation effects.
However, this potential is not easy to dealt with since it has no
periodicity. So it is crucial to write the potential in a tractable
form. In the case of slowly varying deformation, this is done by expanding
the total potential in the first order of strain gradient. Here we
show how this process can be conducted with the help of lattice frame
defined previously. 

In general, the crystal potential is a function of the relative position
between electron and all ions 

\begin{align}
 & \bar{V}(\{\boldsymbol{R}_{\boldsymbol{l}\tau}-\boldsymbol{x}\}),
\end{align}
where $\boldsymbol{x}$ is the position of electrons expressed in
lab frame. $\boldsymbol{l}$ is lattice point label and $\tau$ is
the label of ions inside a unit-cell. Here we assume the position
of ions inside a unit-cell is completely determined by the lattice
points positions while there are exceptions as mentioned earlier in
this paper. Thus $\boldsymbol{R}_{\boldsymbol{l}\tau}(\{\boldsymbol{R}_{\boldsymbol{l}^{\prime}}\})$
can be written as a function of all the lattice points. Due to translational
invariance of the whole crystal, when we displace all lattice points
by the same amount, all ions in a unit-cell will be translated by
the same value as well. This property is described by the formula:
$\boldsymbol{R}_{\boldsymbol{l}\tau}(\{\boldsymbol{R}_{\boldsymbol{l}^{\prime}}\})-\boldsymbol{C}=\boldsymbol{R}_{\boldsymbol{l}\tau}(\{\boldsymbol{R}_{\boldsymbol{l}^{\prime}}-\boldsymbol{C}\})$
with $\boldsymbol{C}$ some constant displacement. Thus we can absorb
the overall constant translation of $\boldsymbol{R}_{\boldsymbol{l}\tau}$
into its $\{\boldsymbol{R}_{\boldsymbol{l}^{\prime}}\}$ dependence.
Thus when $\boldsymbol{C}=\boldsymbol{x}$, the total crystal potential
can be written as a function of the position difference between electron
and lattice points:

\begin{align}
\bar{V}(\{\boldsymbol{R}_{\boldsymbol{l}\tau}-\boldsymbol{x}\})= & \bar{V}(\{\boldsymbol{R}_{\boldsymbol{l}\tau}(\{\boldsymbol{R}_{\boldsymbol{l}^{\prime}}-\boldsymbol{x}\})\})\equiv V(\{\boldsymbol{R}_{\boldsymbol{l}}-\boldsymbol{x}\}),\label{eq:68}
\end{align}
where by defining the crystal potential as $V(\{\boldsymbol{R}_{\boldsymbol{l}}-\boldsymbol{x}\})$
we eliminate the label of ions within a unit-cell. 

The distribution of $\{\boldsymbol{R}_{\boldsymbol{l}}\}$ is not
periodic in general for a deforming crystal. However, for slowly varying
deformation, we can apply local approximation to transform it into
a more tractable form. It is based on the assumption that only ions
within some length scale that is much smaller than the length scale
of strain variation contributes to the above crystal potential. This
is true for metals and non-polar insulators. For polar materials,
the macroscopic electric field caused by polarization need to be attended
to the potential and the argument here applies the local part. With
the lattice field defined in (\ref{eq:1}), we have $\boldsymbol{R}_{\boldsymbol{l}}(t)=\boldsymbol{R}(\boldsymbol{l},t)$.
Expanding $\boldsymbol{l}$ respect to the electron position in lattice
frame $\boldsymbol{x}^{\prime}$, we have: 

\begin{align}
\boldsymbol{R}_{\boldsymbol{l}}(t)-\boldsymbol{x}= & \boldsymbol{R}(\boldsymbol{l},t)-\boldsymbol{R}(\boldsymbol{x}^{\prime},t)\nonumber \\
\approx(l-x^{\prime})^{\alpha}\boldsymbol{a}_{\alpha}(x,t)+ & \frac{1}{2}(l-x^{\prime})^{\alpha}(l-x^{\prime})^{\beta}(\boldsymbol{a}_{\beta}\cdot\boldsymbol{\partial})\boldsymbol{a}_{\alpha}(x,t),\label{eq:18}
\end{align}
where the last term is a first order small quantity proportional to
the spatial gradient of lattice vector fields. Substituting back into
(\ref{eq:68}) and utilizing the property of lattice connection Eq.
(\ref{eq:5}), the Taylor expansion of potential with respect to the
second term in Eq. (\ref{eq:18}) gives the potential as

\begin{align}
V(\{\boldsymbol{R}_{\boldsymbol{l}}-\boldsymbol{x}\}) & \approx V(\{(l^{\alpha}-x^{\prime}{}^{\alpha})\boldsymbol{a}_{\alpha}(\boldsymbol{x},t)\})+\frac{1}{2}\Gamma_{jk}^{i}\hat{O}_{i}^{jk},\label{eq:19}
\end{align}
where $\{\boldsymbol{\Gamma}_{k}\}$ is the lattice connection introduced
in Eq. (\ref{eq:11}). We call the above procedure the gradient expansion
to crystal potential. 

The first term is the potential given by the local lattice at the
electron $\boldsymbol{x}$. It is still not periodic due to the position
dependence of $\boldsymbol{a}_{\alpha}(\boldsymbol{x},t)$. This can
be expected in the lattice bundle picture as electron moves it experiences
different local lattices with potentials of different periodicities.
However, the advantage here is that if we transform to the lattice
frame and apply local approximation, the first term becomes periodic
and tractable. Actually, due to the invariance of crystal potential
under rigid body rotation of all ions and electron at the same time,
we can rotate three lattice vectors freely without changing the crystal
potential i.e $V(\{(l^{\alpha}-x^{\prime}{}^{\alpha})a_{\alpha}^{i}\})=V(\{(l^{\alpha}-x^{\prime}{}^{\alpha})O_{j}^{i}a_{\alpha}^{j}\})$
with $O_{j}^{i}O_{k}^{i}=\delta_{jk}$. This property means that the
first term actually only depends on the lattice frame metric $\{G_{\alpha\beta}=a_{\alpha}^{i}a_{\beta}^{i}\}$. 

The second term is the gradient correction to potential, where $\hat{O}_{i}^{jk}=\sum_{l}\frac{\partial V(\{\tilde{\boldsymbol{R}}_{\boldsymbol{l}}-x^{\prime}{}^{\alpha}\boldsymbol{a}_{\alpha}\})}{\partial\tilde{R}_{l}^{i}}(\tilde{R}_{l}-x^{\prime}{}^{\alpha}a_{\alpha})^{k}(\tilde{R}_{l}-x^{\prime}{}^{\alpha}a_{\alpha})^{j}$
with $\tilde{\boldsymbol{R}}_{\boldsymbol{l}}(\boldsymbol{x},t)=l^{\alpha}\boldsymbol{a}_{\alpha}(\boldsymbol{x},t)$.
It can be understood as the response of crystal potential operator
to strain gradient denoted by $\{\boldsymbol{\Gamma}_{k}\}$ and $\hat{O}_{i}^{jk}$
is the response coefficient. 

\section{Schrodinger equation in lattice frame\label{sec:Schrodinger-equation-in}}

Given the expression of total crystal potential (\ref{eq:19}) up
to first order of strain gradient, the Schrodinger equation in lab
frame reads

\begin{align}
i\partial_{t}\psi= & [-\frac{1}{2m_{e}}\Delta+V_{tot}(\{(l^{\alpha}-x^{\prime}{}^{\alpha})\boldsymbol{a}_{\alpha}\})]\psi,\label{eq:60}
\end{align}
where $\Delta=\partial_{x}^{2}$ is the Laplace operator in lab frame
and

\begin{align}
V_{tot}= & V(\{l^{\alpha}-x^{\prime}{}^{\alpha}\}\boldsymbol{a}_{\alpha})+\frac{1}{2}\Gamma_{jk}^{i}\hat{O}_{i}^{jk}
\end{align}
comes from Eq. (\ref{eq:19}). It is very attempting to express the
Schrodinger equation lattice frame $(\boldsymbol{x}^{\prime},t^{\prime})$
due to the fact that the potential only depends on $\{l^{\alpha}-x^{\prime}{}^{\alpha}\}$
and lattice frame metric. 

During the transformation to lattice frame, if we require that wave-function
invariant under coordinate transformation, then the Schrodinger equation
in lattice frame simply reads

\begin{align}
i\partial_{t^{\prime}}\psi & =\{-\frac{1}{2m}\Delta^{\prime}-iW^{\prime\alpha}\partial_{\alpha}^{\prime}+V_{tot}(\{(l^{\alpha}-x^{\prime}{}^{\alpha})a_{\alpha}\})\}\psi^{\prime},\label{eq:76}
\end{align}
where $\Delta^{\prime}=\frac{1}{\sqrt{G^{\prime}}}\partial^{\prime\alpha}(\sqrt{G}\partial_{\alpha}^{\prime})$
is the Laplacian in lattice frame with $\partial^{\prime\alpha}=G^{\alpha\beta}\partial_{\beta}^{\prime}$,
$\partial_{\beta}^{\prime}=\frac{\partial}{\partial x^{\prime\beta}}$
and $G^{\prime}=det(G_{\alpha\beta})$. $\{G^{\alpha\beta}\}$ is
the inverse matrix of $\{G_{\alpha\beta}\}$ and satisfies $G^{\alpha\beta}G_{\beta\gamma}=\delta_{\gamma}^{\alpha}$,
with repeated indices summed. It has the explicit expression as

\begin{align}
G^{\alpha\beta}= & \frac{\partial x^{\prime\alpha}}{\partial x^{i}}\frac{\partial x^{\prime\beta}}{\partial x^{i}}=b_{i}^{\alpha}b_{i}^{\beta},
\end{align}
where $\{\boldsymbol{b}^{\alpha}\}$ is the reciprocal lattice vector.
In order to be more transparent about the meaning of $G_{0\alpha}$,
we use the symbol $W_{\alpha}^{\prime}=G_{0\alpha}$ to denote this
component of metric tensor. Then we only have spatial indices and
the spatial part of metric tensor $\{G^{\prime\alpha\beta},G_{\alpha\beta}^{\prime}\}$
can be used to raise and lower indices. For example, we have

\begin{align}
W^{\prime\alpha}=G^{\alpha\beta}W_{\beta}^{\prime}=b_{i}^{\alpha}W^{i} & .
\end{align}

However, there is a problem about the Schrodinger equation (\ref{eq:76})
that the Hamiltonian on the right hand side is not Hermitian with
respect to the inner product $\int dx^{\prime}\sqrt{G}$. This inner
product is inherited from the definition in lab frame. Since during
the transformation the wave-function is kept invariant, to ensure
the probability of finding a electron in a given volume is the same
expressed in both coordinates i.e $\Delta\rho=|\psi|^{2}d^{3}x=|\psi|^{2}\sqrt{G}d^{3}x^{\prime}$,
we have to define the inner product in lattice frame as $\int dx^{\prime}\sqrt{G}$.
So to resolve the non-Hermiticity problem, instead of keeping the
wave-function invariant, we require the wave-function in lattice frame
$\psi^{\prime}$ satisfies the following relation

\begin{align}
|\psi|^{2}d^{3}x=|\psi^{\prime}|^{2}d^{3}x^{\prime} & ,\label{eq:61}
\end{align}
as a result of which the physical meaning of wave-function is still
kept while the inner product in lattice frame becomes $\int dx^{\prime}$.
Then we can choose the transformation of wave-function and define
the inner product in lattice frame as 

\begin{align}
\psi^{\prime}\equiv & (G)^{\frac{1}{4}}\psi,\label{eq:69}\\
\langle\varphi^{\prime}|\psi^{\prime}\rangle^{\prime}\equiv & \int\varphi^{\prime*}\psi^{\prime}d^{3}x^{\prime}.\label{eq:81}
\end{align}
As can be seen later, this choice restores the Hermicity of the Hamiltonian
in lattice frame.

To complete the argument, the transformation relation for operators
should also be specified. This can be done by requiring that physical
observables have the same value calculated in both frames:

\begin{align}
 & \int\varphi^{*}\hat{S}\psi d^{3}x=\int\varphi^{\prime*}\hat{S}^{\prime}\psi^{\prime}d^{3}x^{\prime},\label{eq:70}
\end{align}
where $\hat{S}$ and $\hat{S}^{\prime}$ are operators in lab frame
and lattice frame respectively. Thus we see that operators transform
as

\begin{align}
\hat{S}^{\prime}= & |\frac{\partial x}{\partial x^{\prime}}|\hat{S}|\frac{\partial x}{\partial x^{\prime}}|^{-1}.
\end{align}

Equipped with this transformation relation, after some long but tedious
algebra, we finally have the Schrodinger equation for $\psi^{\prime}$
in lattice frame as

\begin{widetext} 

\begin{align}
i\partial_{t^{\prime}}\psi^{\prime}= & [-\frac{1}{2m_{e}}(\partial_{\alpha}^{\prime}-im_{e}W_{\alpha}^{\prime})(G{}^{\prime\alpha\beta}(\partial_{\beta}^{\prime}-im_{e}W_{\beta}^{\prime}))+V_{tot}(\{(l^{\alpha}-x^{\prime}{}^{\alpha})a_{\alpha}(\boldsymbol{x}^{\prime},t^{\prime})\})-\frac{1}{2}m_{e}W_{\alpha}^{\prime}W^{\prime\alpha}+V_{g}]\psi^{\prime},\label{eq:84}
\end{align}

\end{widetext}where $V_{g}=\frac{1}{8m_{e}}\partial_{\alpha}^{\prime}\partial^{\prime\alpha}lnG+\frac{1}{32m_{e}}G^{\alpha\beta}(\partial_{\alpha}^{\prime}lnG)(\partial_{\beta}^{\prime}lnG)$
is a pure geometrical quantity. It is second order in strain gradient
thus will be discarded in our first order theory. It can be checked
that the Hamiltonian in the above Schrodinger equation is Hermitian
with respect to the inner product (\ref{eq:81}).

\section{Lagrangian of wave-packet\label{sec:Lagrangian-of-wave-packet}}

The Schrodinger equation is lattice frame is still hard to solve.
The Hamiltonian in Eq. (\ref{eq:84}) is neither periodic nor static.
However, it is easily seen that both the aperiodicity and time-dependence
come from the fields $\{W^{\prime\alpha}(\boldsymbol{x}^{\prime},t^{\prime}),G^{\alpha\beta}(\boldsymbol{x}^{\prime},t^{\prime})\}$.
So if deformation varies slowly in position and time, local approximation
and adiabatic approximation can be applied to solve this problem.
This can be done systematically with wave-packet method. We refer
to the paper by Sundaram and Niu \cite{sundaram} for a more complete
discussion of this method. The basic idea is that if we have a wave-packet
state of electron that is localized both in real space and reciprocal
space, with its center position in lattice frame as $\boldsymbol{x}_{c}^{\prime}$
and $\boldsymbol{q}_{c}^{\prime}$ respectively, the effective Hamiltonian
is given by the Taylor expansion of position operator in metric fields
relative to the center position of wave-packet. The zeroth order and
first order Hamiltonians for the wave-packet state thus read

\begin{widetext} 

\begin{align}
\hat{\tilde{H}}_{c}^{\prime}= & -\frac{1}{2m}G_{c}^{\alpha\beta}(\partial_{\alpha}^{\prime}-im_{e}W{}_{c\alpha}^{\prime})(\partial_{\beta}^{\prime}-im_{e}W{}_{c\beta}^{\prime})+V(\{(l^{\alpha}-x^{\prime}{}^{\alpha})a_{c\alpha}\})-\frac{1}{2}m_{e}W{}_{c\alpha}^{\prime}W{}_{c}^{\prime\alpha},\label{eq: 69}\\
\Delta\hat{\tilde{H}}_{c}^{\prime}= & \frac{1}{2}[(x{}^{\prime\alpha}-x{}_{c}^{\prime\alpha})\frac{\partial\hat{H}_{c}^{\prime}}{\partial x{}_{c}^{\prime\alpha}}+\frac{\partial\hat{H}_{c}^{\prime}}{\partial x{}_{c}^{\prime\alpha}}(x^{\prime\alpha}-x{}_{c}^{\prime\alpha})]+\frac{1}{2}\Gamma_{jk}^{i}(x_{c}^{\prime},t^{\prime})\hat{O}_{i}^{jk}\{(l^{\alpha}-x^{\prime}{}^{\alpha})a_{c\alpha}\},\label{eq: 71}
\end{align}

\end{widetext}where $\{G_{c}^{\prime\alpha\beta},W_{c\alpha}^{\prime}\}$
are fields evaluated at position $(\boldsymbol{x}_{c}^{\prime},t^{\prime})$. 

The Hamiltonian (\ref{eq: 69}) seems complicated but is actually
easy to solve since it is periodic and the metric tensor and velocity
field are just parameters. To solve this eigen problem, first we define
the gauge invariant wave-vector $k_{c}^{\prime}$ as

\begin{align}
 & \boldsymbol{k}_{c}^{\prime}=\boldsymbol{q}_{c}^{\prime}-m_{e}\boldsymbol{W}_{c}^{\prime}.
\end{align}
Then the eigenstate and eigen value of Hamiltonian (\ref{eq: 69})
reads

\begin{align}
\tilde{\psi}_{c}^{\prime}(\boldsymbol{x}^{\prime};\boldsymbol{q}_{c}^{\prime},\boldsymbol{x}_{c}^{\prime},t^{\prime})= & e^{i\boldsymbol{q}_{c}^{\prime}\boldsymbol{x}^{\prime}}u^{\prime}(\boldsymbol{x}^{\prime};\boldsymbol{k}_{c}^{\prime},\boldsymbol{x}_{c}^{\prime},t^{\prime}),\label{eq: 73}\\
\tilde{\varepsilon}_{c}^{\prime}(\boldsymbol{q}_{c}^{\prime},\boldsymbol{x}_{c}^{\prime},t^{\prime})= & \varepsilon_{c}^{\prime}(\boldsymbol{k}_{c}^{\prime},\boldsymbol{x}_{c}^{\prime},t^{\prime})-\frac{1}{2}m_{e}W{}_{c}^{\prime\alpha}W{}_{c\alpha}^{\prime},\label{eq: 74}
\end{align}
where $u^{\prime}(\boldsymbol{x}^{\prime};\boldsymbol{k}_{c}^{\prime},\boldsymbol{x}_{c}^{\prime},t^{\prime})$
and $\varepsilon_{c}^{\prime}(\boldsymbol{k}_{c}^{\prime},\boldsymbol{x}_{c}^{\prime},t^{\prime})$
are eigen-states and eigen-energy of the Hamiltonian without velocity
field
\begin{align}
\hat{H}_{c}^{\prime}(\boldsymbol{k}_{c}^{\prime},\boldsymbol{x}_{c}^{\prime},t^{\prime})= & -\frac{1}{2m}G_{c}^{\alpha\beta}(\partial_{\alpha}^{\prime}+ik{}_{c\alpha}^{\prime})(\partial_{\beta}^{\prime}+ik_{c\beta}^{\prime})\nonumber \\
 & +V(\{(l^{\alpha}-x^{\prime}{}^{\alpha})\boldsymbol{a}_{c\alpha}\}).\label{eq: 75}
\end{align}

As a first order theory in inhomogeneity, we do not need to consider
the correction to the wave-function from the first order Hamiltonian
(\ref{eq: 71}). But we do need to consider its correction to energy.
To calculate this gradient correction, we first superpose the eigenstates
(\ref{eq: 73}) to construct a electron wave-packet state. Then calculate
the expectation value of the first order Hamiltonian (\ref{eq: 71})
in this wave-packet state. Again this process is quite standard in
wave-packet method \cite{sundaram}, we just list the result here

\begin{align}
\Delta\tilde{\varepsilon}_{c}^{\prime}(\boldsymbol{q}_{c}^{\prime},\boldsymbol{x}_{c}^{\prime},t^{\prime})= & -\operatorname{Im}\langle\partial_{\boldsymbol{x}_{c}^{\prime}}|_{\boldsymbol{q}_{c}^{\prime}}u^{\prime}\mid\cdot(\varepsilon_{c}^{\prime}-\hat{H}_{c}^{\prime})\mid\partial_{\boldsymbol{q}_{c}^{\prime}}u^{\prime}\rangle^{\prime}\nonumber \\
 & +\frac{1}{2}\Gamma_{cjk}^{i}\langle u^{\prime}\mid\hat{O}_{i}^{jk}\mid u^{\prime}\rangle^{\prime},
\end{align}
where the first and second terms come from the two terms in (\ref{eq: 71})
respectively. After expressing $(\partial_{\boldsymbol{q}_{c}^{\prime}},\partial_{\boldsymbol{x}_{c}^{\prime}})$
in terms of $(\partial_{\boldsymbol{k}_{c}^{\prime}},\partial_{\boldsymbol{x}_{c}^{\prime}})$,
the first term gives rise to two terms: $2\boldsymbol{\omega}_{c}^{\prime}\cdot\boldsymbol{J}^{\prime}$
and $\boldsymbol{Im}\langle\partial_{\boldsymbol{x}_{c}^{\prime}}|_{\boldsymbol{k}_{c}^{\prime}}u^{\prime}\mid\cdot(\varepsilon_{c}^{\prime}-\hat{H}_{c}^{\prime})\mid\partial_{\boldsymbol{k}_{c}^{\prime}}u^{\prime}\rangle^{\prime}$.
$\boldsymbol{\omega}_{c}^{\prime}$ defined as $\boldsymbol{\omega}_{c}^{\prime}=\frac{1}{2}\partial_{\boldsymbol{x}_{c}^{\prime}}\times\boldsymbol{W}_{c}^{\prime}$
is the augular velocity of lattice and $\boldsymbol{J}^{\prime}=\frac{m_{e}}{2}\operatorname{Im}\langle\partial_{\boldsymbol{k}_{c}^{\prime}}u^{\prime}\mid\times(\varepsilon_{c}^{\prime}-\hat{H}_{c}^{\prime})\mid\partial_{\boldsymbol{k}_{c}^{\prime}}u^{\prime}\rangle^{\prime}$
is the angular momentum of electron. 

Then the Lagrangian for wave-packet in lattice frame reads

\begin{widetext} 

\begin{align}
L_{e}^{\prime}= & -(\varepsilon_{tot}^{\prime}-\frac{1}{2}m_{e}W{}_{c}^{\prime\alpha}W{}_{c\alpha}^{\prime})+(\boldsymbol{k}_{c}^{\prime}+m_{e}\boldsymbol{W}_{c}^{\prime})\cdot\dot{\boldsymbol{x}}_{c}^{\prime}+\langle\tilde{u}^{\prime}|i\partial_{\boldsymbol{x}_{c}^{\prime}}\tilde{u}^{\prime}\rangle^{\prime}\cdot\dot{\boldsymbol{x}}_{c}^{\prime}+\langle u^{\prime}|i\partial_{\boldsymbol{k}_{c}^{\prime}}u^{\prime}\rangle^{\prime}\cdot\dot{\boldsymbol{k}}_{c}^{\prime}+\langle u^{\prime}|i\partial_{t^{\prime}}u^{\prime}\rangle^{\prime},\label{eq: 77}
\end{align}

\end{widetext}with $\varepsilon_{tot}^{\prime}=\varepsilon_{c}^{\prime}-\frac{1}{2}m_{e}W{}_{c}^{\prime\alpha}W{}_{c\alpha}^{\prime}+2\boldsymbol{\omega}_{c}^{\prime}\cdot\boldsymbol{J}^{\prime}+\frac{1}{2}\boldsymbol{\Gamma}_{c}\langle\boldsymbol{O}\rangle^{\prime}+\operatorname{Im}\langle\partial_{\boldsymbol{x}_{c}^{\prime}}u^{\prime}\mid\cdot(\varepsilon_{c}^{\prime}-\hat{H}_{c}^{\prime})\mid\partial_{\boldsymbol{k}_{c}^{\prime}}u^{\prime}\rangle^{\prime}$
is the energy depending on the Bloch functions. 

The Lagrangian of wave-packet in lattice frame (\ref{eq: 77}) is
very useful in deriving results but its physical meaning is usually
less clear since we are more accustomed to understand a physical problem
in lab frame. Also we are left with the question how to calculate
the eigen problem of Hamiltonian (\ref{eq: 75}) with Ab initial calculations.
Further more, lattice frame is not globally well-defined in the presence
of defects so we are unable to consider the topological effect associated
with defects in lattice frame. Based on the above reasons, it is more
desirable to express the results in lab frame.

First, we rewrite the Hamiltonian (\ref{eq: 75}) in an orthonormal
coordinates defined as

\begin{align}
r^{i}= & x^{\prime}{}^{\alpha}a_{c\alpha}^{i}.\label{eq:96}
\end{align}
Then the Hamiltonian written in this orthonormal coordinates denoted
by $\hat{H}_{c}$ reads

\begin{align}
\hat{H}_{c}(k_{c};x_{c},t)=\frac{1}{2m_{e}}(\frac{1}{i}\frac{\partial}{\partial\boldsymbol{r}}+\boldsymbol{k}_{c})^{2}+V(\{l^{\alpha}\boldsymbol{a}_{c\alpha}(\boldsymbol{x_{c}},t)-\boldsymbol{r}\}) & ,\label{eq: 79}
\end{align}
which is exactly the Hamiltonian (\ref{7}) evaluated at $\boldsymbol{x_{c}}$
with $k_{ci}\equiv k_{c\alpha}^{\prime}b_{ic}^{\alpha}$. Thus we
automatically get the transformation rule for wave-vector from lattice
frame to lab frame 

\begin{align}
k_{ci}= & k_{c\alpha}^{\prime}b_{i}^{\alpha}(\boldsymbol{x_{c}},t),
\end{align}
which denotes points in the Brillouin zone of the local lattice at
$(\boldsymbol{x_{c}},t)$.

Then the eigen solution of (\ref{eq: 75}) is related to the eigen
solution of (\ref{eq: 79}) as

\begin{align}
u^{\prime}(\boldsymbol{x}^{\prime};\boldsymbol{k}_{c}^{\prime},\boldsymbol{x}_{c}^{\prime},t^{\prime})= & u(x^{\prime}{}^{\alpha}\boldsymbol{a}_{c\alpha};k_{c\alpha}^{\prime}\boldsymbol{b}_{c}^{\alpha},\boldsymbol{x}_{c}(\boldsymbol{x}_{c}^{\prime},t),t),\\
\varepsilon_{c}^{\prime}(\boldsymbol{k}_{c}^{\prime},\boldsymbol{x}_{c}^{\prime},t^{\prime})= & \varepsilon_{c}(k_{c\alpha}^{\prime}\boldsymbol{b}_{c}^{\alpha},\boldsymbol{x}_{c}(\boldsymbol{x}_{c}^{\prime},t),t),
\end{align}
where $u(\boldsymbol{r};\boldsymbol{k}_{c},\boldsymbol{x}_{c},t)$
and $\varepsilon_{c}(\boldsymbol{k}_{c},\boldsymbol{x}_{c},t)$ are
the eigen states and eigen energy of Hamiltonian (\ref{eq: 79}).
Then the Berry connection in the Lagrangian (\ref{eq: 77}) take the
form

\begin{align}
\langle u^{\prime}|i\partial_{x_{c}^{\prime\mu}}u^{\prime}\rangle^{\prime}= & \frac{\partial x_{c}^{\nu}}{\partial x_{c}^{\prime\mu}}[\frac{(2\pi)^{3}}{v_{c}(\boldsymbol{x}_{c},t)}\int d^{3}ru^{*}i\nabla_{x_{c}^{\nu}}u],
\end{align}
where $v_{c}(\boldsymbol{x}_{c},t)$ is the unit-cell volume at point
$(\boldsymbol{x}_{c},t)$ and the lattice covariant derivative $\nabla_{x_{c}^{\nu}}$
arises naturally. Thus through this line of derivation, we have proved
the validity of lattice covariant derivative. 

Thus if we define the Berry connection in lab frame as 

\begin{align}
A_{x_{c}^{\nu}}\equiv i\langle u|\nabla_{x_{c}^{\nu}}|u\rangle\equiv & \frac{(2\pi)^{3}}{v}\int d^{3}ru^{*}\nabla_{x_{c}^{\nu}}u,\\
A_{k_{ci}}\equiv i\langle u|\partial_{k_{ci}}|u\rangle\equiv & \frac{(2\pi)^{3}}{v}\int d^{3}ru^{*}\partial_{k_{ci}}u,
\end{align}
then the relation between Berry connections in lattice frame and lab
frame reads $A_{x_{c}^{\prime\mu}}=\frac{\partial x_{c}^{\nu}}{\partial x_{c}^{\prime\mu}}A_{x_{c}^{\nu}}$
and $A_{k_{c\alpha}^{\prime}}^{\prime}=\frac{\partial x_{c}^{\prime\alpha}}{\partial x_{c}^{j}}A_{k_{cj}}$. 

The next step is to transform the Lagrangian back to lab frame. Since
time is the same for both coordinates, their Lagrangians can also
be chosen to be equal for a particular gauge. Then we can write down
the Lagrangian in lab frame as

\begin{align}
L_{e}= & -(\varepsilon_{tot}-\frac{1}{2}m_{e}\boldsymbol{W}_{c}^{2})+(\boldsymbol{k}_{c}+m_{e}\boldsymbol{W}_{c})(\dot{x}_{c}-\boldsymbol{W}_{c})\nonumber \\
 & +(A_{t}+\dot{\boldsymbol{x}}_{c}\cdot A_{\boldsymbol{x}_{c}})+D_{t}\boldsymbol{k}_{c}\cdot A_{\boldsymbol{k}_{c}},\label{eq:83}
\end{align}
where

\begin{align}
\varepsilon_{tot}= & \varepsilon_{c}+\boldsymbol{\Gamma}_{c}\operatorname{Im}\langle u|\boldsymbol{D}-\hat{\boldsymbol{D}}|\partial_{k}u\rangle+\boldsymbol{\Gamma}_{c}\langle\hat{\boldsymbol{O}}\rangle+2\boldsymbol{\omega}_{c}\cdot\boldsymbol{J}.
\end{align}
The first term is the eigen energy of local Hamiltonian (\ref{7})
evaluated at $\boldsymbol{x}_{c}$. The second and third term come
from the gradient correction to local Hamiltonian Eq. (\ref{eq:19-1}).
The last term is due to lattice motion rotation with time and $\boldsymbol{J}$
is the angular momentum of electrons. After omitting the indices $c$
we have the Lagrangian appearing in the main text (\ref{eq:83-1}).

\section{Orbital magnetization and polarization\label{sec:Orbital-magnetization-and}}

Here, we deduce the result in Eq. (\ref{eq:49-1}). For a insulator,
the total current density up to first order is given by two contributions

\begin{align}
j_{a}^{i}= & \int\dot{x}^{i}Dd^{3}k,\\
j_{b}^{i}= & \partial_{j}\int\frac{dk}{(2\pi)^{3}}Im[\langle\frac{\partial u_{nk}}{\partial k_{i}}|H-\varepsilon|\frac{\partial u_{nk}}{\partial k_{j}}\rangle],
\end{align}
where $D=1+tr(\Omega_{\boldsymbol{kx}})-m_{e}2\boldsymbol{\omega}\cdot\Omega_{\boldsymbol{k}}$
is the density of state. $\boldsymbol{j}_{b}$ comes from the dipole
moments of velocity operator \cite{xiao2010berry}. We can combine
Eq. (\ref{eq:51}) and (\ref{eq:52}) to solve for $\dot{\boldsymbol{x}}$. 

Especially the last two terms in Eq. (\ref{eq:49-1}) come from the
following terms in $\boldsymbol{j}_{a}$:

\begin{align}
 & e\int d^{3}k[-\Omega_{kx}\cdot W+\Omega_{kt}+W\cdot tr(\Omega_{kx})].\label{eq:92}
\end{align}
To get the same form in Eq. (\ref{eq:49-1}), we will frequently use
the following identity 

\begin{align}
 & \partial_{x^{\mu}}\int dkf(k,x,t)=-\Gamma_{i\mu}^{i}\int dkf+\int dk\nabla_{x^{\mu}}f,\label{eq:101}
\end{align}
where $f$ is any phase space function. Choosing the periodic gauge
$A_{x^{\mu}}(k+2\pi b;x,t)=A_{x^{\mu}}(k;x,t)$ and noticing that
$\nabla_{x^{j}}A_{k_{j}}=\partial_{x^{j}}A_{k_{j}}-\Gamma_{mj}^{n}k_{n}\partial_{k_{m}}A_{k_{j}}-\Gamma_{ij}^{j}A_{k_{i}}$,
we have

\begin{align}
\int-\Omega_{k_{i}x^{j}}W^{j}= & W^{j}(\partial_{j}P^{i}-\Gamma_{lj}^{i}P^{l}+\Gamma_{lj}^{l}P^{i}),\\
\int\Omega_{k_{i}t}= & \partial_{t}P^{i}-\Gamma_{l0}^{i}P^{l}+\Gamma_{l0}^{l}P^{i},\\
\int W^{i}tr(\Omega_{kx})= & -W^{i}(\partial_{j}P^{j}-\Gamma_{lj}^{j}P^{l}+\Gamma_{lj}^{l}P^{j}),
\end{align}
where $\boldsymbol{P}=-\frac{e}{(2\pi)^{3}}\int A_{\boldsymbol{k}}$.
Substituting back to Eq. (\ref{eq:92}) and utilizing the elastic
condition $\Gamma_{i0}^{k}+\Gamma_{ji}^{k}W^{j}=\partial_{i}W^{k}$,
we have the last two terms in Eq. (\ref{eq:49-1}):

\begin{align}
 & \partial_{t}\boldsymbol{P}-\boldsymbol{\partial}\times(\boldsymbol{W}\times\boldsymbol{P}).\label{eq:58-2}
\end{align}

Next, we deduce the Chern-Simons contribution to flexoelectricity
and dynamical magnetization. We start from the expressions in lattice
frame then transform back to lab frame. Because lattice connection
vanishes in lattice frame, the results in \cite{inhomogeneouspolarization}
can be applied directly in lattice frame. We can directly write down
the Chern-Simons contribution to current density as

\begin{align}
j_{cs}^{i}= & -e\int dk^{\prime}[\Omega_{k_{i}^{\prime}k_{j}^{\prime}}^{\prime}\Omega_{x_{j}^{\prime}t^{\prime}}^{\prime}+\Omega_{k_{j}^{\prime}x^{\prime j}}^{\prime}\Omega_{k_{i}^{\prime}t^{\prime}}^{\prime}+\Omega_{x_{j}^{\prime}k_{i}^{\prime}}^{\prime}\Omega_{k_{j}^{\prime}t^{\prime}}^{\prime}]\nonumber \\
= & e\partial_{t^{\prime}}\int dk^{\prime}\partial_{k_{i}^{\prime}}A_{k_{j}^{\prime}}^{\prime}A_{x_{j}^{\prime}}^{\prime}+\partial_{k_{j}^{\prime}}A_{x_{j}^{\prime}}^{\prime}A_{k_{i}^{\prime}}^{\prime}+\partial_{x_{j}^{\prime}}A_{k_{i}^{\prime}}^{\prime}A_{k_{j}^{\prime}}^{\prime}\nonumber \\
 & -e\partial_{x_{j}^{\prime}}\int dk^{\prime}\partial_{k_{i}^{\prime}}A_{k_{j}^{\prime}}^{\prime}A_{t}^{\prime}+\partial_{k_{j}}A_{t}^{\prime}A_{k_{i}}^{\prime}+\partial_{t}^{\prime}A_{k_{i}}^{\prime}A_{k_{j}}^{\prime}.\label{eq:95}
\end{align}
where the periodic gauge condition for real space Berry connections
has been used:

\begin{align*}
A_{t^{\prime}}^{\prime}(\boldsymbol{k}^{\prime})-A_{t^{\prime}}^{\prime}(\boldsymbol{k}^{\prime}+2\pi\boldsymbol{b}^{\prime}) & =0,\\
A_{\boldsymbol{x}^{\prime}}^{\prime}(\boldsymbol{k}^{\prime})-A_{\boldsymbol{x}^{\prime}}^{\prime}(\boldsymbol{k}^{\prime}+2\pi\boldsymbol{b}^{\prime}) & =0,
\end{align*}
where $\boldsymbol{b}^{\prime}$ is the reciprocal lattice vector
in lattice frame. From (\ref{eq:95}), we can easily identify the
polarization and magnetization term as

\begin{align}
P^{\prime i}= & e\int dk^{\prime}A_{k_{i}^{\prime}}\partial_{x^{\prime j}}A_{k_{j}^{\prime}}+A_{k_{j}^{\prime}}\partial_{k_{i}^{\prime}}A_{x^{\prime j}}+A_{x^{\prime j}}\partial_{k_{j}^{\prime}}A_{k_{i}^{\prime}},\\
M^{\prime ij}= & -e\int dk^{\prime}\partial_{k_{i}^{\prime}}A_{k_{j}^{\prime}}A_{t^{\prime}}+\partial_{k_{j}^{\prime}}A_{t^{\prime}}A_{k_{i}^{\prime}}+\partial_{t^{\prime}}A_{k_{i}^{\prime}}A_{k_{j}^{\prime}}.
\end{align}
Then using the transformation rule of polarization-magnetization tensor
under coordinate transformation, which we have verified in zeroth
order, we can get the results given by Eq. (\ref{eq:53-1}) and (\ref{eq:54-3}).

\section{Energy stress tensor of electron\label{sec:Energy-stress-tensor}}

Here we will demonstrate how the concept of energy stress tensor appears
naturally by considering the role electrons play in the dynamics of
ions. From Appendix \ref{sec:Lattice-frame}, we can see the fundamental
field describing the ionic degrees of freedom is contained in the
lattice field $\boldsymbol{R}(\boldsymbol{x}^{\prime},t^{\prime})$
given by Eq. (\ref{eq:1}). If we are to consider the dynamics of
ions in the least action principle, we need to variate both electron
and ion action with respect to this lattice field. Here, we focus
the electron part and see how the electron energy stress tensor emerges. 

The electron Lagrangian written in lattice frame depends the of lattice
frame metric field, which are related to lattice field as

\begin{align}
G_{\alpha\beta}= & \partial_{\alpha}R^{i}\partial_{\beta}R^{i}=a_{\alpha}^{i}a_{\beta}^{i},\\
G_{0\alpha}= & \partial_{\alpha}R^{i}\partial_{t^{\prime}}R^{i}=a_{\alpha}^{i}W^{i}.
\end{align}
\begin{widetext} 

We then write the action of electron wave-packet in the following
form:

\begin{align}
A= & \int_{V_{0}}d\boldsymbol{x}^{\prime}dt^{\prime}\delta(\boldsymbol{x}^{\prime}-\boldsymbol{y}^{\prime}(t^{\prime}))L_{e}^{\prime}(W^{\prime}(\boldsymbol{y}^{\prime},t^{\prime}),\{G_{\alpha\beta}^{\prime}(\boldsymbol{y}^{\prime},t^{\prime})\};(\boldsymbol{k}^{\prime}(t^{\prime}),\dot{\boldsymbol{y}}^{\prime}(t^{\prime}),\dot{\boldsymbol{k}}^{\prime}(t^{\prime}))),
\end{align}
\end{widetext}where $L_{e}^{\prime}$ is given by Eq. (\ref{eq: 77}).
Here we use variables $(\dot{\boldsymbol{k}}^{\prime}(t^{\prime}),\dot{\boldsymbol{y}}^{\prime}(t^{\prime}),\boldsymbol{y}^{\prime}(t^{\prime}),t^{\prime})$
to denote the degrees of freedom of electron as a point particle,
all of which are functions of time $t^{\prime}$. And we add a factor
$\delta(\boldsymbol{x}^{\prime}-\boldsymbol{y}^{\prime}(t^{\prime}))$
to express the action in a field form so that we can apply the variation
principle in field theories directly. 

Then we variate the above action with respect to the lattice field
$\boldsymbol{R}(\boldsymbol{x}^{\prime},t^{\prime})$. After some
long and tedious calculation, we have

\begin{widetext} 
\begin{align}
F^{i}=\frac{\delta S_{e}}{\delta R^{i}}(x^{\prime},t^{\prime})= & \frac{\partial L_{e}}{\partial G_{\alpha\beta}}[\partial_{y^{\prime\alpha}}\delta(\boldsymbol{x}^{\prime}-\boldsymbol{y}^{\prime})\partial_{y^{\prime\beta}}R^{i}+\alpha\leftrightarrow\beta]|_{t^{\prime}}+\frac{\partial L_{e}}{\partial(\partial_{\gamma}G_{\alpha\beta})}\partial_{y^{\prime\gamma}}[\partial_{y^{\prime\alpha}}\delta(\boldsymbol{x}^{\prime}-\boldsymbol{y}^{\prime})\partial_{y^{\prime\beta}}R^{i}+\alpha\leftrightarrow\beta]\nonumber \\
 & -\{\tilde{\partial}_{t^{\prime}}[\frac{\partial L_{e}}{\partial(\partial_{t^{\prime}}G_{\alpha\beta})}]\partial_{y^{\prime\beta}}R^{i}\partial_{y^{\prime\alpha}}\delta(\boldsymbol{x}^{\prime}-\boldsymbol{y}^{\prime})+\alpha\leftrightarrow\beta\}\nonumber \\
 & +[\frac{\partial L_{e}}{\partial G_{0\alpha}}\partial_{y^{\prime\alpha}}\delta(\boldsymbol{x}^{\prime}-\boldsymbol{y}^{\prime})\partial_{t^{\prime}}R^{i}-\delta(\boldsymbol{x}^{\prime}-\boldsymbol{y}^{\prime})\tilde{\partial}_{t^{\prime}}(\frac{\partial L_{e}}{\partial G_{0\alpha}}\partial_{y^{\prime\alpha}}R^{i})]\nonumber \\
 & +\frac{\partial L_{e}}{\partial(\partial_{\gamma}G_{0\alpha})}\partial_{y^{\prime\gamma}}[\partial_{y^{\prime\alpha}}\delta(\boldsymbol{x}^{\prime}-\boldsymbol{y}^{\prime})\partial_{t^{\prime}}R^{i}]-\tilde{\partial}_{t^{\prime}}\{\frac{\partial L_{e}}{\partial(\partial_{\gamma}G_{0\alpha})}\partial_{y^{\prime\gamma}}[\delta(\boldsymbol{x}^{\prime}-\boldsymbol{y}^{\prime})\partial_{y^{\prime\alpha}}R^{i}\}\label{eq: 115}
\end{align}
\end{widetext}where the derivative operator $\tilde{\partial}_{t^{\prime}}$
only acts the explicit time dependence not on those variables $(\dot{\boldsymbol{k}}^{\prime}(t^{\prime}),\dot{\boldsymbol{y}}^{\prime}(t^{\prime}),\boldsymbol{y}^{\prime}(t^{\prime}))$.
Here we still haven't used the expressions of $(\dot{\boldsymbol{k}}^{\prime},\dot{\boldsymbol{y}}^{\prime})$
. $\frac{\delta S_{e}}{\delta R^{j}}$ describes the force density
electron exerting on ions. We see from the delta function $\delta(\boldsymbol{x}^{\prime}-\boldsymbol{y}^{\prime})$
that the effective interaction between electron and ion is local,
which is inherited from the local approximation and adiabatic approximation
we used. 

Eq. (\ref{eq: 115}) is the contribution from a single electron. In
reality, we have multiple electrons filling in the band structure.
For simplicity, here we only consider a particular band in a insulator
at zero temperature. Then we need to sum over all electrons in a filled
band with the integration of density of state and substitute the expression
for $(\dot{\boldsymbol{k}}^{\prime},\dot{\boldsymbol{y}}^{\prime})$
given by the equations of motion. Then the total force from all electron
contributions reads

\begin{align}
\mathcal{F}^{i}(\boldsymbol{x}^{\prime},t^{\prime})= & \int dy^{\prime}dk^{\prime}D^{\prime}\frac{\delta A_{e}}{\delta R^{i}}(\boldsymbol{x}^{\prime},t^{\prime}),
\end{align}
where we adopt the convention to use mathcal form of a symbol to denote
all electron contributions such as $\mathcal{F}$ here. 

Then substituting Eq. (\ref{eq: 115}) into the above expression,
we have

\begin{align}
\mathcal{F}^{i}(x^{\prime},t^{\prime})= & \sqrt{G}[\partial_{j}\mathcal{T}^{ij}+\partial_{t}\mathcal{T}^{0i}],\label{eq: 117}
\end{align}
where

\begin{align}
\mathcal{T}^{ij}= & \frac{\partial x^{i}}{\partial x^{\prime\alpha}}\frac{\partial x^{j}}{\partial x^{\prime\beta}}\mathcal{T}^{\alpha\beta}+\mathcal{T}^{0i}W^{j}+W^{i}\mathcal{T}^{0j},\label{eq: 118}\\
\mathcal{T}^{0i}= & \frac{\partial x^{i}}{\partial x^{\prime\beta}}T^{0\beta}\label{eq: 119}
\end{align}
and $\mathcal{T}^{\alpha\beta},\mathcal{T}^{0\beta}$ are the four-dimensional
energy stress tensor of electrons in lattice frame defined as: 

\begin{widetext} 

\begin{align}
\mathcal{T}^{\alpha\beta}(\boldsymbol{x}^{\prime},t^{\prime})= & -\int dk^{\prime}D^{\prime}\frac{2}{\sqrt{G}}\{\frac{\tilde{\partial}L}{\tilde{\partial}G_{\alpha\beta}}-\partial_{\sigma}\frac{\tilde{\partial}L_{e}}{\tilde{\partial}(\partial_{\sigma}G_{\alpha\beta})}-\tilde{\partial}_{t^{\prime}}\frac{\tilde{\partial}L_{e}}{\tilde{\partial}(\partial_{t^{\prime}}G_{\gamma\delta})}\},\label{eq: 120}\\
\mathcal{T}^{0\alpha}(\boldsymbol{x}^{\prime},t^{\prime})= & -\int dk^{\prime}D^{\prime}\frac{1}{\sqrt{G}}\{\frac{\tilde{\partial}L_{e}}{\tilde{\partial}G_{0\alpha}}-\partial_{\sigma}\frac{\tilde{\partial}L_{e}}{\tilde{\partial}(\partial_{\sigma}G_{0\alpha})}\},\label{eq:121}
\end{align}

\end{widetext} where $\frac{\tilde{\partial}}{\tilde{\partial}G_{\alpha\beta}},\frac{\tilde{\partial}}{\tilde{\partial}(\partial_{\sigma}G_{\alpha\beta})},\frac{\tilde{\partial}}{\tilde{\partial}(\partial_{t^{\prime}}G_{\gamma\delta})},\frac{\tilde{\partial}}{\tilde{\partial}G_{0\alpha}},\frac{\tilde{\partial}}{\tilde{\partial}(\partial_{\sigma}G_{0\alpha})}$
do not act on the metric dependence of $(\dot{\boldsymbol{k}}^{\prime},\dot{\boldsymbol{y}}^{\prime})$
whose expression is given by the equations of motion and $D^{\prime}=(1+tr(\Omega_{k^{\prime}x^{\prime}}^{\prime})-2m_{e}\boldsymbol{\omega}^{\prime}\cdot\boldsymbol{\Omega}^{\prime})$

Here we come up with the expression (\ref{eq: 120}, \ref{eq:121})
to ensure the physical meaning of energy stress tensor i.e the form
of Eq. (\ref{eq: 117}). Equivalently, we can define the energy stress
tensor directly in lattice frame following the variation to metric
as

\begin{align}
T^{\alpha\beta}\equiv & -2\frac{1}{\sqrt{G}}\frac{\delta A}{\delta G_{\alpha\beta}},\label{eq: 122}\\
T^{0\alpha}\equiv & -\frac{1}{\sqrt{G}}\frac{\delta A}{\delta G_{0\alpha}},\label{eq: 123}
\end{align}
where the definitions of $T^{\alpha\beta}$ and $T^{0\alpha}$ differ
by a factor of two besides their variations to different components
of metric, which is a feature of non-relativistic theory \cite{NCgeometry}.
Then following the same procedure in calculating $\mathcal{F}^{i}$
from $F^{i}$, we can achieve the expressions (\ref{eq: 120}, \ref{eq:121}).

To calculate the specific form of energy stress tensor, we need to
substitute the expression of electron Lagrangian Eq. (\ref{eq: 77})
into Eq. (\ref{eq: 120}, \ref{eq:121}). For simplicity, here we
consider the special case where strain is homogeneous in space but
varies with time such that the metric tensor is only time dependent
then $D^{\prime}=(1-2m_{e}\boldsymbol{\omega}^{\prime}\cdot\boldsymbol{\Omega}^{\prime})$.
Then we have $L_{e}^{\prime}=-(\varepsilon^{\prime}-\frac{1}{2}m_{e}G{}^{0\gamma}G{}_{0\gamma}+2\boldsymbol{\omega}^{\prime}\cdot\boldsymbol{J}^{\prime})+(\boldsymbol{k}^{\prime}+m_{e}\boldsymbol{W}^{\prime})\cdot\dot{\boldsymbol{y}}^{\prime}+A_{\boldsymbol{k}^{\prime}}^{\prime}\cdot\dot{\boldsymbol{k}}^{\prime}+\partial_{t^{\prime}}G_{\sigma\gamma}\langle u^{\prime}|i\tilde{\partial}_{G_{\sigma\gamma}}u^{\prime}\rangle^{\prime}$.
Then following the definition (\ref{eq: 120}), variation to $G_{\alpha\beta}$
while keeping $G_{0\alpha}=W_{\alpha}^{\prime}$ fixed gives 

\begin{widetext} 

\begin{align*}
T^{\alpha\beta}= & \frac{2}{\sqrt{G}}\int dk^{\prime}D^{\prime}\{\frac{\tilde{\partial}\varepsilon^{\prime}}{\tilde{\partial}G_{\alpha\beta}}+\frac{1}{2}m_{e}G{}^{0\alpha}G{}^{0\beta}\\
 & +m_{e}\partial_{\sigma}G_{0\gamma}\operatorname{Im}\tilde{\partial}_{G_{\alpha\beta}}\langle\partial_{k_{\gamma}^{\prime}}u^{\prime}\mid(\varepsilon^{\prime}-\hat{H}^{\prime})\mid\partial_{k_{\sigma}^{\prime}}u^{\prime}\rangle^{\prime}-\dot{\boldsymbol{k}}^{\prime}\cdot\frac{\tilde{\partial}A_{\boldsymbol{k}^{\prime}}^{\prime}}{\tilde{\partial}G_{\alpha\beta}}-\partial_{t^{\prime}}G_{\sigma\gamma}\Omega_{G_{\alpha\beta}G_{\sigma\gamma}}\}\\
= & 2\int\frac{dk^{\prime}}{\sqrt{G}}\{\tilde{\partial}_{G_{\alpha\beta}}\varepsilon^{\prime}+\frac{1}{2}m_{e}G^{0\alpha}G{}^{0\beta}\\
 & +m_{e}\partial_{\sigma}G_{0\gamma}\operatorname{Im}\tilde{\partial}_{G_{\alpha\beta}}\langle\partial_{k_{\sigma}^{\prime}}u^{\prime}\mid(\varepsilon^{\prime}+\hat{H}^{\prime})\mid\partial_{k_{\gamma}^{\prime}}u^{\prime}\rangle^{\prime}+m_{e}a_{\gamma}^{l}\partial_{t^{\prime}}W^{l}\tilde{\partial}_{G_{\alpha\beta}}A_{k_{\gamma}^{\prime}}^{\prime}-\partial_{t^{\prime}}G_{\sigma\gamma}\Omega_{G_{\alpha\beta}G_{\sigma\gamma}}\},
\end{align*}

\end{widetext} where from the second line to third line, we substitute
the expression of $\dot{\boldsymbol{k}}^{\prime}$ and use the periodic
gauge condition for $\langle u^{\prime}|i\partial_{G_{\alpha\beta}}u^{\prime}\rangle^{\prime}$
. $\Omega_{G_{\alpha\beta}G_{\sigma\gamma}}=i[\partial_{G_{\alpha\beta}}\langle u^{\prime}|\partial_{G_{\sigma\gamma}}u^{\prime}\rangle^{\prime}-\partial_{G_{\sigma\gamma}}\langle u^{\prime}|\partial_{G_{\alpha\beta}}u^{\prime}\rangle^{\prime}]$
is the Berry curvature with respect to lattice metric. On the other
hand,

\begin{align*}
T^{0\alpha}= & -\int dk^{\prime}D^{\prime}\frac{1}{\sqrt{G}}\{\frac{\tilde{\partial}L_{e}}{\tilde{\partial}G_{0\alpha}}\}\\
= & -\int\frac{dk^{\prime}}{\sqrt{G}}D^{\prime}\{m_{e}W{}^{\prime\alpha}+m_{e}\dot{x}^{\prime\alpha}]\}
\end{align*}
Then following Eq. (\ref{eq: 118}, \ref{eq: 119}) and using the
identity $2\frac{\partial}{\partial G_{\alpha\beta}}=b_{i}^{\beta}\frac{\partial}{\partial a_{\alpha}^{i}}$,
we have the energy stress tensor in lab frame as 

\begin{align}
\mathcal{T}_{j}^{i}= & \mathcal{D}_{j}^{i}-2\boldsymbol{\omega}\cdot\nabla_{j}^{i}\boldsymbol{\mathcal{J}}+\nabla_{j}^{i}P_{m_{e}}\cdot a-\Gamma_{l0}^{m}\int dk\Omega_{im}^{jl}\nonumber \\
 & +\frac{m_{e}}{v}W{}^{i}W{}^{j}-[W^{j}j_{m_{e}}^{i}+i\leftrightarrow j]
\end{align}
We will illustrate the physical meaning of each term in the main body
of this paper.

\section{Expressions in momentum representation\label{sec:Expressions-in-momentum}}

Although we introduce the concept of lattice covariant in the position
representation in order to contrast with the normal partial derivatives,
often it is more convenient to calculate in momentum space. So it
is worth to discuss the expression in momentum space. Since the local
Hamiltonians and Bloch states all have the same periodicity as local
lattices, their expression in momentum space only involves discrete
momentum basis $|\boldsymbol{l}\rangle=exp(2\pi il_{\alpha}\boldsymbol{b}^{\alpha}(\boldsymbol{x},t)\cdot\boldsymbol{r})$,
where $\boldsymbol{l}$ is some integer and $\boldsymbol{b}(\boldsymbol{x},t)$
the reciprocal lattice vector at $(\boldsymbol{x},t)$. If we calculate
the lattice covariant derivative of Bloch states expressed on the
momentum basis, we have
\begin{align}
\nabla_{x^{\mu}}u(\boldsymbol{r},\boldsymbol{k};\boldsymbol{x},t)= & \sum_{\boldsymbol{l}}[\partial_{x^{\mu}}u_{\boldsymbol{l}}+\Gamma_{m\mu}^{n}k_{n}\partial_{k_{m}}u_{\boldsymbol{l}}]exp(il_{\alpha}\boldsymbol{b}^{\alpha}\cdot\boldsymbol{r}),
\end{align}
where $u_{\boldsymbol{l}}(\boldsymbol{k};\boldsymbol{x},t)=\int d\boldsymbol{r}u(\boldsymbol{r},\boldsymbol{k};\boldsymbol{x},t)exp(-2\pi il_{\alpha}\boldsymbol{b}^{\alpha}(\boldsymbol{x},t)\cdot\boldsymbol{r})$
is the Fourier component of Bloch functions. An important property
is that the basis $|\boldsymbol{l}\rangle$ vanishes under lattice
covariant derivative while not under partial derivative. Thus in the
momentum representation lattice covariant derivative takes a simpler
form as

\begin{align}
\nabla_{x^{\mu}}= & \partial_{x^{\mu}}+\Gamma_{m\mu}^{n}k_{n}\partial_{k_{m}},\label{eq:56}
\end{align}
which acting on the Fourier components. Then the real space Berry
connection can be conveniently expressed as
\begin{align}
A_{x^{\mu}}= & i\Gamma_{m\mu}^{n}\sum_{\boldsymbol{l}}u_{\boldsymbol{l}}^{*}(a_{\alpha}^{m}\partial_{a_{\alpha}^{n}}+k_{n}\partial_{k_{m}})u_{\boldsymbol{l}}.\label{eq:58-1}
\end{align}
where the relation $\partial_{x^{\mu}}=\Gamma_{m\mu}^{n}a_{\alpha}^{m}\partial_{a_{\alpha}^{n}}$
is used. The normalization condition for Bloch function is $\sum_{\boldsymbol{l}}u_{\boldsymbol{l}}^{*}u_{\boldsymbol{l}}=1$
from Eq. (\ref{eq:41}). 

For the local Hamiltonian, its matrix element expressed in momentum
representation is
\begin{align}
H_{l,l^{\prime}}= & \frac{(\boldsymbol{G}_{\boldsymbol{l}}+\boldsymbol{k})^{2}}{2m}\delta_{\boldsymbol{l},\boldsymbol{l}^{\prime}}+V(\boldsymbol{G}_{\boldsymbol{l}-\boldsymbol{l}^{\prime}}(\boldsymbol{x},t)),
\end{align}
where $\boldsymbol{G}_{\boldsymbol{l}}=2\pi l_{\alpha}\boldsymbol{b}^{\alpha}$,
$\boldsymbol{G}_{\boldsymbol{l}-\boldsymbol{l}^{\prime}}=2\pi(l-l^{\prime})_{\alpha}\boldsymbol{b}^{\alpha}$
and $V(\boldsymbol{G})=\frac{1}{v}\int drV(\{\tilde{R}_{l}-r\})exp(i\boldsymbol{G}\cdot\boldsymbol{r})$
is the Fourier component of the local crystal potential. Directly
applying (\ref{eq:56}), we have the lattice covariant derivative
of local Hamiltonian as
\begin{align}
\nabla_{x^{\mu}}\hat{H}_{l,l^{\prime}}= & \hat{D}_{n;l,l^{\prime}}^{m}\Gamma_{m\mu}^{n},
\end{align}
where $\{\hat{D}_{n;l,l^{\prime}}^{m}\}$ is the deformation potential
operator in momentum representation, which reads

\begin{align}
\hat{D}_{n;l,l^{\prime}}^{m}= & -[\frac{(G_{l}+k)_{m}(G_{l}+k)_{n}}{2m}\delta_{l,l^{\prime}}+\nabla_{n}^{m}V(\boldsymbol{G}_{\boldsymbol{l}-\boldsymbol{l}^{\prime}})].
\end{align}
$\nabla_{n}^{m}V(\boldsymbol{G})=\lim_{\delta G_{l}\rightarrow0}\frac{V(\boldsymbol{G}+\delta\boldsymbol{G})-V(\boldsymbol{G})}{\delta G_{m}^{\alpha}}G_{n}^{\alpha}$
is the derivative of crystal potential to strain expressed in reciprocal
space. $V(\boldsymbol{G}+\delta\boldsymbol{G})$ and $V(\boldsymbol{G})$
correspond to two crystals with different periodicities. The deformation
potential expressed in reciprocal space has been discussed in \cite{deformationpotential}.
However, in that paper the term given by (2.14b) shouldn't appear.
Because during Fourier transformation, the fact that the integral
domain also changes after deformation is not considered. Also, we
prove here that the result is valid in general case not limited to
the rigid ion model. 

The energy effect of deformation operator has been discussed thoroughly
\cite{herring1956transport}, which is the diagonal part in the Bloch
basis. Its off-diagonal part also plays an important role in electron
dynamics through Berry curvatures. This can be seen through the expression
of Berry curvatures as a sum over all bands contributions. For example,
$\Omega_{nk_{i}}^{m}$ can be written as as

\begin{align}
\Omega_{nk_{i}}^{m}= & i\sum_{l\neq0}\frac{\langle u_{0}|\hat{D}_{n}^{m}|u_{l}\rangle\langle u_{l}|\hat{v}_{i}|u_{0}\rangle-\langle u_{0}|\hat{v}_{i}|u_{l}\rangle\langle u_{l}|\hat{D}_{n}^{m}|u_{0}\rangle}{(\varepsilon_{0}-\varepsilon_{l})^{2}},\label{eq:125}
\end{align}
where $0$ is the band we are interested and $l$ labels all the other
bands. The role of deformation potential operator is explicit in this
expression. Integratation in Brillouin zone $-e\int d^{3}ki\sum_{l\neq0}\frac{\langle0|\hat{O}_{mn}|l\rangle\langle l|\hat{v}_{i}|0\rangle-\langle0|\hat{v}_{i}|l\rangle\langle l|\hat{O}_{mn}|0\rangle}{(\varepsilon_{0}-\varepsilon_{l})^{2}}$
just gives the proper piezoelectric constant. 

And viscosity term comes from the Berry curvature $\Omega_{nq}^{mp}$
involving strain and can be written as

\begin{widetext}

\begin{align}
\Omega_{nq}^{mp}=i & \sum_{l\neq0}\frac{\langle u_{0}|\hat{D}_{n}^{m}|u_{l}\rangle\langle u_{l}|\hat{D}_{q}^{p}|u_{0}\rangle-\langle u_{0}|\hat{D}_{q}^{p}|u_{l}\rangle\langle u_{l}|\hat{D}_{n}^{m}|u_{0}\rangle}{(\varepsilon_{0}-\varepsilon_{l})^{2}}.\label{eq:125-1}
\end{align}

\end{widetext}

\end{document}